\documentclass[prd,amsmath,amssymb,superscriptaddress,floatfix,nofootinbib,notitlepage]{revtex4-1}

\usepackage{bbold}
\usepackage{color}
\usepackage{graphicx}
\usepackage{hyperref}

\newcommand{\Eq}[1]{{Eq.~({\ref{#1}})}}

\newcommand{\Fig}[1]{{Fig.~{\ref{#1}}}}

\newcommand{\bea}{\begin{eqnarray}}
\newcommand{\eea}{\end{eqnarray}}
\newcommand{\be}{\begin{equation}}
\newcommand{\ee}{\end{equation}}
\newcommand{\beas}{\begin{eqnarray*}}
\newcommand{\eeas}{\end{eqnarray*}}

\newcommand{\nn}{\nonumber}

\def\k{{\bf k}}
\def\r{{\bf r}}

\def\als{{\alpha_s}}
\def \cf {C_F}

\def \aver {{\langle r \rangle_{nl}}}
\def \averr {{\langle r^2 \rangle_{nl}}}
\def \averrz {{\langle r^2 \rangle_{10}}}

\begin{document}

\title{Heavy Quarkonium at finite temperature and chemical potential}

\author{Stefano~Carignano}
\author{Joan Soto}
\affiliation{Departament de F\'isica Qu\`antica i Astrof\'isica and Institut de Ci\`encies del Cosmos, Universitat de Barcelona, Mart\'i i Franqu\`es 1, 08028 Barcelona, Catalonia, Spain.}

\begin{abstract}
  We generalize known results for heavy quarkonium in a thermal bath to the case of a finite baryonic density, and 
   provide a number of formulas for the energy shift and decay width that hold at weak coupling for sufficiently large temperature and/or chemical potential.
We find that a non-vanishing decay width requires a temperature larger than the typical binding energy, no matter how large the chemical potential is. This implies that at zero temperature the dissociation mechanism of heavy quarkonium is due entirely to screening, unlike in the finite temperature case. We use several effective theories in order to sort out the contributions of the relevant energy and momentum scales. In particular, we consider contributions of the so called quasi-static magnetic modes. The generalization to the case of a finite isospin/strangeness chemical potential is trivial. We discuss possible applications to the SIS and NICA conditions, and compare with available lattice results.

\end{abstract}

\maketitle

%\tableofcontents

\section{Introduction}

High energy heavy-ion collision experiments (HIC) have shown the existence of collective behavior in the strong interactions, namely  a new state of matter that is usually refer to as quark-gluon plasma (QGP) (see \cite{Braun-Munzinger:2015hba} for a review). In order to study the properties of the QGP, the so called hard probes have been extremely useful. Among them, the suppression of heavy quarkonium states in the products of the HIC were proposed as a signal of QGP formation long ago \cite{Matsui:1986dk} (see \cite{Aarts:2016hap,Andronic:2015wma,Rothkopf:2019ipj} for reviews). Nowadays the sequential suppression of $\Upsilon (1,2,3)$ has been clearly observed \cite{Sirunyan:2018nsz,Acharya:2018mni}. However, the QCD dynamics which is actually responsible for this suppression is not so easy to identify. The original proposal of Matsui and Satz \cite{Matsui:1986dk} that screening would be the mechanism behind sequential melting of quarkonium bound states is not entirely correct. In \cite{Laine:2006ns}, it was shown that for a weakly interacting QGP the same dynamics that produces screening, also produces an imaginary part to the potential, as a consequence of the so called Landau damping. In \cite{Escobedo:2008sy}, it was emphasized that this imaginary part is parametrically larger than the real part
and hence Landau damping rather than screening should be regarded as the key mechanism for heavy quarkonium dissociation.
Imaginary potentials were also obtained in strongly coupled QGP settings \cite{Laine:2007qy,Burnier:2015nsa,Burnier:2014ssa} and included in models of in-medium heavy quarkonium \cite{Margotta:2011ta,Dumitru:2009fy,Boyd:2019arx,Guo:2018vwy}.
Later on, within the weakly coupled QGP, detailed analysis were made taking into account the interplay between temperature, screening mass and the different scales in the quarkonium bound state dynamics, the main lesson being that finite temperature effects cannot always be incorporated in a phenomenological potential \cite{Escobedo:2008sy,Brambilla:2008cx,Escobedo:2010tu,Brambilla:2010vq}. The effects of a relative velocity of the heavy quarkonium with respect to the medium have also been analysed \cite{Escobedo:2011ie,Escobedo:2013tca} (see also \cite{Aarts:2012ka}). Recently, part of these findings have been embedded in a more realistic framework of an expanding QGP, no matter whether this is weakly or strongly coupled \cite{Brambilla:2016wgg,Brambilla:2017zei,Blaizot:2017ypk,Yao:2018nmy,Yao:2018sgn,Blaizot:2018oev,Brambilla:2019tpt,Eller:2019spw}.

High energy HIC experiments are essentially gluon colliders, and the resulting medium has a negligible baryonic chemical potential with respect to the temperature scales attained. In the near future, however, there are planned HIC experiments at lower energies that aim at attaining large values of the chemical potential at SIS (CBM) \cite{Friman:2011zz}  and Nica (MPD) \cite{Sissakian:2009zza} (see \cite{Galatyuk:2019lcf} for a recent overview). These colliders will have energy enough to produce charmonium bound states \cite{Ablyazimov:2017guv}. 
It is then worth exploring in a solid theoretical framework, namely using QCD at weak coupling and the well-known effective field theories for heavy quarkonium, the fate of these states at non-zero baryon chemical potential. This is so even if the charm quark mass may not be high enough to apply weak coupling techniques beyond the ground state, or if the values of 
chemical potential actually attained in the experiments may not be large enough to justify a weak coupling analysis. Indeed, the weak coupling analysis may unravel qualitative new features that may then be incorporated in more realistic models or settings. This was the case, for instance, when an imaginary part of the potential at finite temperature was uncovered
in \cite{Laine:2006ns}. 

We shall then restrict ourselves to the study of a heavy quarkonium propagating in a QGP in thermodynamical equilibrium  such that the temperatures $T$ and baryon chemical potentials $\mu$ fulfill $m \gg T \gg g\,T \gg \Lambda_{QCD}$ and $m \gg \mu \gg g\mu \gg \Lambda_{QCD}$, where $g$ is the QCD coupling constant and $m$ is the heavy quark mass. 
We shall also assume that the heavy quarkonium is weakly coupled, namely $m\gg  m\als \gg m\als^2 \gtrsim  \Lambda_{QCD}$,
where $\als=g^2/4\pi$, and is at rest in the QGP rest frame. Recall that $p\sim m\als$ is the typical heavy quark momentum in the bound state (and $r\sim 1/m\als$ its typical radius) and $E\sim m\als^2$ the typical size of the binding energy. 
We aim at the calculation of the leading order effects both in $\mu$ and in $T$ on the mass (binding energy) and decay width. 
These calculations are non-trivial because on the one hand at energy scales of the order of or below the typical quarkonium binding energy Coulomb
ressummations must be carried out, and on the other hand at momentum scales of the order of or below the Debye mass HTL resummations must be carried out. In that respect, the use of suitable effective field theories \cite{Caswell:1985ui,Braaten:1991gm,Pineda:1997bj} is very convenient.

We shall distribute the paper as follows. In Sec. \ref{bf} we set the basic formalism, in Sec. \ref{pTmuE} and \ref{TmupE} we address the two most significant cases, $p\gg {\rm max}(T\,,\mu) \gg E$ and ${\rm max}(T\,,\mu)\gg p \gg E$, respectively.
 Each section contains subsections where  the cases $T\gtrsim \mu$ and $\mu \gg T$ are separately addressed. 
 Sec. \ref{disc} discusses our results as well as other cases that are not addressed in full detail. The more technical developments are relegated to the appendices.

\section{Basic formalism}
\label{bf}

Throughout this work we will use the real-time formalism for thermal field theory (for reviews see eg. \cite{Thoma:2000dc,Laine:2016hma,Ghiglieri:2020dpq}), including both the effects of temperature and chemical potential, following the lines of \cite{Brambilla:2008cx,Brambilla:2010vq}. 
We shall consider the heavy quarks and heavy quarkonium as probe particles, and hence absent in the medium. This means in practice that  
the real-time non-relativistic propagator reduces for them to the $11$ components only, and those take the form of the usual non-relativistic retarded propagators at zero temperature and chemical potential.

In the real-time formalism, the longitudinal and transverse  gluon propagators are four-by-four matrices diagonal in color space which at tree level in the Coulomb gauge read (color indices omitted), respectively 
\cite{Landshoff:1992ne} 
\bea
 {\bf D}^{(0)}_{00}({\bf k}) &=&
\left(
\begin{matrix}
&&\hspace{-2mm} \displaystyle \frac{i}{{k}^2}
&&0
\\
&&\hspace{-2mm} 0
&&\displaystyle -\frac{i}{k^2}
\end{matrix}
\right),
\label{D000}
\\
 {\bf D}^{(0)}_{ij}(K) &=&
\left(\delta_{ij} - \frac{k^ik^j}{k^2}\right)
\!\left\{\!
\left(
\begin{matrix}
&&\hspace{-2mm} \displaystyle \frac{i}{K^2 + i\epsilon}
&&\theta(-k^0) \, 2\pi\delta(K^2)
\\
&&\hspace{-2mm} \theta(k^0) \, 2\pi\delta(K^2)
&&\displaystyle -\frac{i}{K^2 - i\epsilon}
\end{matrix}
\right)
+ 2\pi\delta(K^2)\, n_{\rm B}(|k^0|)\,
\begin{pmatrix}
1 & 1 \\
1 & 1
\end{pmatrix}
\right\}\!,
\nn\\
\label{D0ij}
\eea
where $K = (k_0,{\k})$ and  $k = |\k|$.
Note that the longitudinal part of the gluon propagator in Coulomb gauge does not depend on the
temperature.  
We can write them in terms of advanced/retarded propagators,
\be
[D_{\mu\nu}]_{11}= \frac{ D_{\mu\nu}^{\rm R}(k_0,k) + D_{\mu\nu}^{\rm A}(k_0,k)}{2}
    + \left[\frac{1}{2} + n_{\rm B}(k_0)\right]\left(D_{\mu\nu}^{\rm R}(k_0,k) - D_{\mu\nu}^{\rm A}(k_0,k)\right) \,,
  \label{eq:11comp}
\ee
$n_{\rm B}$ being the (temperature-dependent) bosonic occupation number. 
This formula holds at any order. At tree level,
\be
D_{ij}^{\rm R\,,A}(k^0,\k)=i\frac{\delta_{ij} - k_i k_j/k^2}{K^2 \pm ik^0 \epsilon}\quad ,\quad D_{00}^{\rm R\,,A}(k^0,\k)=\frac{i}{k^2 } \,,\qquad D^{\rm R\,,A}_{0i}=D^{\rm R\,,A}_{i0}=0 \,,
\ee
where ``R'' stands for retarded and ``A'' for advanced.
 Note that the property 
\be
 D_{\mu\nu}^{\rm R\,,A}(-k^0,\k)=D_{\mu\nu}^{\rm A\,,R}(k^0,\k)\quad ,\quad [D_{\mu\nu}]_{11}(-k^0,\k)=[D_{\mu\nu}]_{11}(k^0,\k)\,
 \label{eq:symmD}
\ee
is inherited by the full propagators, and will be often used in the following.

When integrating out the hard scale, namely for $K \ll $ max$(T,\mu)$, the gauge sector reduces to that of the well-established Hard Thermal Loop (HTL) % HTL
effective theory, whose longitudinal and transverse propagators read
\begin{equation}
D^{\mathrm{R,A}}_{00}(k_0,k)=
\frac{i}{k^2+m_D^2\left(1-\displaystyle\frac{k_0}{2k}\log\frac{k_0+k\pm i\eta}{k_0-k\pm i\eta}\right)}\, \qquad {\rm and} 
\label{prophtllong}
\end{equation}
\begin{equation}
D^{\mathrm{R,A}}_{ij}(k_0,k)=\left(\delta_{ij}-\frac{k_ik_j}{k^2}\right)\Delta_\mathrm{R,A}(k_0,k)\,,
\label{prophtltrans}
\end{equation}
respectively,  where
\begin{equation}
\Delta_\mathrm{R,A}(k_0,k)=
\frac{i}{k_0^2-k^2-\displaystyle\frac{m_D^2}{2}
\left(\displaystyle\frac{k_0^2}{k^2}-(k_0^2-k^2)\displaystyle\frac{k_0}{2k^3}
\log\left(\displaystyle\frac{k_0+k\pm i\eta}{k_0-k\pm i\eta}\right)\right)\pm i\,\mathrm{sgn}(k_0)\,\eta}\,.
\label{defdelta}
\end{equation}

For a QCD medium at finite temperature and density with $N_f$ light quark flavors the expression for the Debye mass $m_D$ reads
\cite{Vija:1994is} 
\be
m_D^2 = g^2 \Big[ 
T_F N_f \left( \frac{T^2}{3} + \frac{\mu^2}{\pi^2} \right) + N_c \frac{T^2}{3} \Big] \equiv m_{D(F)}^2 + m_{D(B)}^2 \,,
\label{eq:mD}
\ee
where $N_c$ is the number of colors, $T_F = 1/2$ and we have separated the bosonic ($m_{D(B)}^2 \sim N_c$) and fermionic ($m_{D(F)}^2  \sim N_f$) contribution to it for later use.

At energy and momentum scales smaller than the Debye mass $m_D$, the longitudinal gluons are screened and can be integrated out.
However, a particular class of transverse gluons, the so called quasi-static magnetic modes, survive at those lower scales. They fulfill $m_D\gg k \gg k_0$. Hence, in this case the transverse propagator can be approximated by
\be
\Delta_\mathrm{R,A}(k_0,k) \simeq \Delta_\mathrm{R,A}^{(M)}(k_0,k)=
\frac{i}{-k^2\pm\frac{i\,\pi m_D k_0}{4 k}} \,.
\label{magnetic} 
\ee

Having introduced the formalism we will use throughout the paper, we now move on to investigate the effects of a dense medium on quarkonium states.

\section{The case $m \gg p \gg {\rm max}(T,\mu) \gg E$ } 
\label{pTmuE}

We start by considering, and extending to finite chemical potential, the case discussed in
\cite{Escobedo:2010tu,Brambilla:2010vq,Brambilla:2017zei}, namely 
$m \gg p \gg {\rm max}(T,\mu) \gg E $.
In this case, the $T$ and $\mu$ will affect the binding energy and decay width of the quarkonium, but not its size. The heavy quarkonium essentially remains a Coulombic bound state, and the medium effects are perturbations to it. Since scales much larger than ${\rm max}(T,\mu)$ lead to exponentially suppressed Boltzmann factors, we can start our considerations directly from the 
 (vacuum) pNRQCD Lagrangian \cite{Pineda:1997bj,Brambilla:1999xf}, namely
\begin{eqnarray}
{\cal L}_{\textrm{pNRQCD}} &=
&
- \frac{1}{4} F^a_{\mu \nu} F^{a\,\mu \nu}
+ \sum_{i=1}^{N_f}\bar{q}_i\,iD\!\!\!\!/\,q_i
+ \int d^3r \; {\rm Tr} \,
\Biggl\{ {\rm S}^\dagger \left[ i\partial_0 - h_s \right] {\rm S}
+ {\rm O}^\dagger \left[ iD_0 -h_o \right] {\rm O} \Biggr\}
\nonumber\\
&& \hspace{-1.5cm}
+ V_A\, {\rm Tr} \left\{  {\rm O}^\dagger \r \cdot g{\bf E} \,{\rm S}
+ {\rm S}^\dagger \r \cdot g{\bf E} \,{\rm O} \right\}
+ \frac{V_B}{2} {\rm Tr} \left\{  {\rm O}^\dagger \r\cdot g{\bf E} \, {\rm O}
+ {\rm O}^\dagger {\rm O} \r \cdot g{\bf E}  \right\}  + \dots\,,
\label{pNRQCD}
\end{eqnarray}
with $E^i = F^{i0}$ chromo-electric field. The singlet/octet Hamiltonians are
\begin{equation}
h_{s/o}=\frac{\mathbf{p}^2}{m}+\frac{\mathbf{P}^2}{4m}
+V^{(0)}_{s/o}
+\frac{V^{(1)}_{s/o}}{m}+\frac{V^{(2)}_{s/o}}{m^2}+\ldots,
\label{sinoctham}
\end{equation}
where $\mathbf{P}$ and $\mathbf{p}$ are the center-of-mass and relative momentum respectively, and the various $V^{(n)}$ are potentials known up to a certain order. In our calculations, $\mathbf{P}$, $V_B$ and the subleading potentials ($n>0$) can be neglected and 
we may approximate $V^{(0)}_{s}\simeq -C_F\als/r$, $V^{(0)}_{o}\simeq (N_c/2-C_F)\als/r$, and $V_A\simeq 1$.

Now we may integrate out the largest scale, $T$ or $\mu$, and get to another EFT which is valid at the lower scales  $E,m_D$.  
 The outcome will be a new contribution to the singlet potential: $ V_s \to V_s + \delta V$, with 

\be
\delta V=- i g^2 \, C_F \, \frac{r^i}{D-1}
\nu^{4-D} \int \frac{d^Dk}{(2\pi)^D}
\frac{i}{E-h_o-k_0 +i\eta}\Big(k_0^2 \,  [D^{}_{ii}(k_0,k)]_{11} +  k^2 \,  [D^{}_{00}(k_0,k)]_{11} \Big) r^i\,,
\label{eq:deltav1}
\ee
where $D^{}_{\mu\nu}(k_0,k)$ stands for the full gluon propagator in the Coulomb gauge and we are using dimensional regularization (DR) with $D= d+1 = 4+2\epsilon$, $\nu$  the DR subtraction scale, and
\begin{equation}
\int {d^Dk} = \int_{-\infty}^{\infty} dk^0 \int d\Omega_d \int_0^\infty dk k^{d-1} \,,
\end{equation}
$\Omega_d$ denoting the solid angle in $d$ spatial dimensions.

As mentioned in the previous section, quarkonium and heavy quarks are considered in this work as test particles outside of the medium. 
As a consequence, in the real-time formalism only the $11$ components of the gluon propagators, $[D_{\mu\nu}]_{11}$, will couple to them. 
In the following, we will omit for brevity the 11 indices and only label explicitly the retarded and advanced components when they appear.

\subsection{Integrating out the hard scale }
\label{hard}

Integrating out the hard scale ($T$ or $\mu$) will give us a new EFT which we will refer to as pNRQCD$_{HTL}$, following 
\cite{Brambilla:2010vq}.
In addition, if the hard scale is much larger than $ E\sim h_o$, we can expand the octet propagator\footnote{In principle there can be a region $k_0\sim E$, $k\sim$ max$(T,\mu)$, that should also be integrated out, for which this expansion does not hold. In the case $T\sim \mu$, it leads to subleading contributions.},
\be
\frac{i}{E-h_o-k_0+i\eta}=
\frac{i}{-k_0+i\eta}-i\frac{E-h_o}{(-k_0+i\eta)^2}
+i\frac{(E-h_o)^2}{(-k_0+i\eta)^3} + \dots \,. 
\label{octetexpand}
\ee

\begin{figure}{
\includegraphics[width=.3\textwidth]{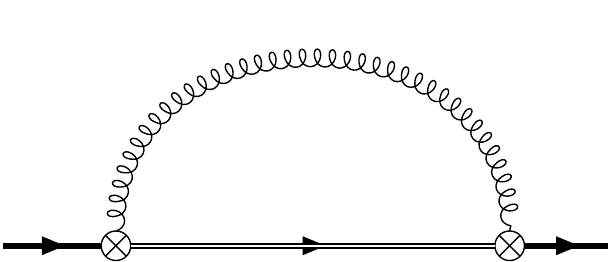}
\caption{One-loop contribution to the quarkonium self-energy. The solid thick lines denote the quarkonium singlet propagator and the double line the quarkonium octet propagator. Curly lines denote transverse gluon propagators and crossed dots are chromoelectric dipole vertices.  \label{fig:self1}}}
\end{figure}

From the general expression \Eq{eq:deltav1}, we will consider two contributions:

\subsubsection{One-loop hard contribution (\Fig{fig:self1}).} 

In this case, $D^{}_{\mu\nu}(k_0,k)=D^{(0)}_{\mu\nu}(k_0,k)$ in \Eq{D000} and \Eq{D0ij}. 
Since the tree level longitudinal propagator does not depend on the distribution function, the last term in \Eq{eq:deltav1} can be dropped. Moreover, since $D^{(0)}_{\mu\nu}(k_0,k)=D^{(0)}_{\mu\nu}(-k_0,k)$, the first term in the expansion \Eq{octetexpand} leads to a vanishing contribution. The contribution from the second term 
has been computed in 
\cite{Brambilla:2008cx,Brambilla:2010vq}. It  does not contain any fermionic occupation number - the only medium 
dependence enters in the $n_B$ from the $11$ gluon propagator prescription. So there will not be any $\mu$ dependence, and we can just take those results.
One obtains 
\begin{align}
\delta V
& =\frac{\pi}{9}N_c C_F \als^2 T^2r+\frac{2\pi}{3m}C_F\als T^2  + {\cal O}(\frac{\als E^2}{m}) \,.
\label{deltaVlinearpluscubic}
\end{align} 
Since the contribution from the first term of \Eq{octetexpand} vanishes for symmetry reasons and \Eq{deltaVlinearpluscubic} comes from the second-order term in the expansion, there could be higher-loop diagrams that give a contribution comparable to it.
We analyze them in the next section.

\begin{figure}{
\includegraphics[width=.3\textwidth]{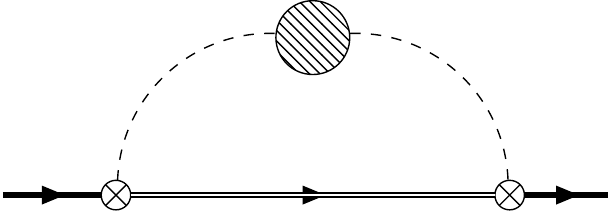}
\caption{Two-loop contribution to the quarkonium self-energy. The solid thick lines denote the quarkonium singlet propagator and the double line the quarkonium octet propagator. Dashed lines denote the longitudinal gluon propagator, crossed dots are chromoelectric dipole vertices and the blob denotes the longitudinal gluon self-energy.
 \label{fig:self2}}}
\end{figure}

\subsubsection{The two-loop hard contribution (\Fig{fig:self2}). }\label{2lh}

At two loops, the longitudinal gluons may now contribute because the one-loop self-energy provides them with a $T$ and $\mu$ dependence. The transverse gluons give subleading contributions because the would-be leading term in \Eq{octetexpand} vanishes for the same symmetry reasons as in the previous section. Hence, \Eq{eq:deltav1} reduces to,
\be
\delta V = - i g^2 \, C_F \, \frac{r^i}{D-1}
\nu^{4-D} \int \frac{d^Dk}{(2\pi)^D}
\frac{i}{-k_0 +i\eta}
 k^2 \,  [D^{}_{00}(k_0,k)] \,
r^i\,,
\label{eq:defleading}
\ee
where $D^{}_{00}(k_0,k)$ must be calculated at one loop.
Moreover,  we can write
\be
\frac{i}{-k_0 \pm i\eta} = - i {\cal {P}}(1/k^0) \pm \pi \delta(-k^0) \,,
\label{eq:Pplusdelta}
\ee
where ${\cal{P}}$ denotes the principal value integral. Using again the symmetry properties of $D$, \Eq{eq:symmD}, we see that only the delta function piece survives. Hence, as long as we work at the lowest order  of the expansion \Eq{octetexpand}, the symmetry of the problem forces $k^0 \to 0$ and
 we only need to calculate $\Pi_{00}(k^0\to 0,k)$.

\begin{figure}[ht]
\includegraphics[width=.4\textwidth]{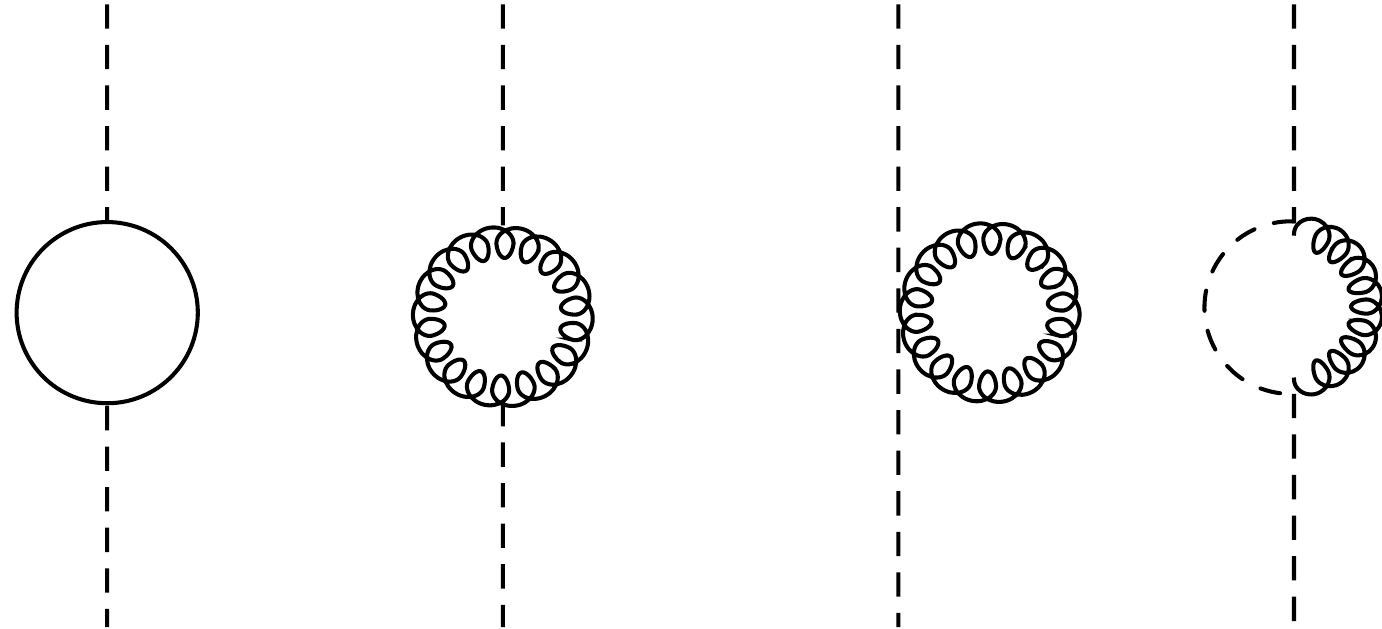}
\caption{Diagrams contributing to the longitudinal component of the
gluon polarization tensor at one-loop order (taken from \cite{Brambilla:2008cx}). The solid line stands for the
light (massless) quark propagator, the dashed line for the longitudinal gluon propagator and
the curly line for the transverse gluon propagator.
Ghosts do not contribute to the thermal part of the gluon polarization tensor
\cite{Landshoff:1992ne}.}
\label{figvacpol}
\end{figure}

For the one-loop longitudinal gluon self-energy $\Pi_{00}(k_0,k)$ we have to sum the gluonic and fermionic loop contributions shown in  Fig.~\ref{figvacpol}.
With our definitions we can write
\bea
\Pi_{00}(k^0,\k)
&=& \Pi_{00,\,{\rm F}}(k^0,\k)
+ \Pi_{00,\,{\rm G}}(k^0,\k)\,,
\label{Pi00full} 
\eea
where  ``F'' labels the
contribution coming from the loops of $N_f$ massless quarks (first diagram of Fig.~\ref{figvacpol})
and ``G'' labels the contribution from the second, third and fourth diagram of Fig.~\ref{figvacpol}.

The gluon contribution to the self-energy is unchanged by the presence of a chemical potential and can be taken directly from \cite{Brambilla:2008cx}. 
Our focus will then be on the fermionic contribution, which at $\mu=0$ and finite $T$ is given by \cite{Brambilla:2008cx},
\begin{align}
\Pi_{00,\,{\rm F}}(k^0,\k)
&=
\frac{g^2 \, T_F \, N_f}{2\pi^2}\int_{-\infty}^{+\infty}dq^0\,|q^0|\, n_{\rm F}(\vert q^0\vert)
\times\Bigg[2-\left(\frac{4q_0^2+k_0^2 - k^2-4q^0k^0}{4|q^0| k}\right)
\log\Big(\frac{k_0^2 -k^2 -2q^0k^0+2|q^0| k}{k_0^2 - k^2 -2q^0k^0-2|q^0| k}\Big) \nn \\
& +\left(\frac{4q_0^2+k_0^2 - k^2+4q^0k^0}{4|q^0| k}\right)
\log\Big(\frac{k_0^2 - k^2 +2q^0k^0-2|q^0| k}{k_0^2 - k^2 +2q^0k^0+2|q^0| k }\Big)\Bigg]\,.
\label{Pi00fermion}
\end{align}
Note that $\Pi_{00,\,{\rm F}}(-k^0,\k)=\Pi_{00,\,{\rm F}}(k^0,\k)$.
We need to generalize this fermionic contribution to finite $\mu$. Recall that in our real-time Feynman rules (see eg. \cite{Carignano:2019ofj}) the occupation number enters as a sgn$(q^0) N_{B/F}(q^0)$, 
with $N_{B/F}(q^0)=1 \pm 2 n_{B/F}(q^0)$, $N_{B/F}(-q^0)=-N_{B/F}(q^0)$.  For $\mu=0$ this reduces to $1\pm 2 n_{B/F}(|q^0|)$.
In a dense medium the fermionic occupation number instead is given by $n_F(q^0-\mu)$ and 
 we face expressions like 
\be
\int dq^0  \delta(Q^2) {\rm sgn}(q^0) N_F(q^0 -\mu) f(q^0)=
\frac{1}{2q} \big[ N_F(q - \mu) f(q) + N_F(q + \mu)f(-q) \big]  \,,
\label{eq:rulemu}
\ee
and if $f(-q^0)=f(q^0)$, as in \Eq{Pi00fermion},
 we can just replace in our expressions
\be
2 n_F(q) \rightarrow   n_F(q - \mu) + n_F(q + \mu)  \qquad (q>0)\,.
\ee

Furthermore, from the discussion after \Eq{eq:Pplusdelta}, we know that we only need the small $k^0$ limit 
of the longitudinal gluon self-energy.
If we expand $\Pi_{00}^{\rm R}(k)$
and $\Pi_{00}^{\rm A}(k)$ for $k^0 \ll k$
and keep terms up to order $k^0$,
the result for its real and imaginary parts is
\bea
{\cal R}_\Pi \equiv {\rm Re} \, \left[\Pi_{00}^{\rm R}(k_0\to 0, k)\right]=
{\rm Re} \, \left[\Pi_{00}^{\rm A}(k_0\to 0,k)\right] &=&
\nn\\
&& \hspace{-7cm}
\frac{g^2 \, T_F \, N_f}{\pi^2}
\int_0^{+\infty}dq\,q\,
\left( \frac{n_{\rm F}(q-\mu)+n_{\rm F}(q+\mu)}{2} \right)
\left[
2 +  \left( \frac{k}{2q} - 2 \frac{q}{k} \right)
\log \left| \frac{k-2q}{k+2q} \right|
\right]
\nn\\
&& \hspace{-7cm}
+ \frac{g^2 \, N_c}{\pi^2}
\int_0^{+\infty}dq\,q\,n_{\rm B}(q)\left[
1 - \frac{k^2}{2q^2}
+ \left( - \frac{q}{k} + \frac{k}{2q} - \frac{k^3}{8q^3} \right)
\log \left| \frac{k-2q}{k+2q} \right|
\right] \,,
\nn\\
\label{RePi00k0}\\
{\cal I}_\Pi  \equiv {\rm Im} \, \left[\Pi_{00}^{\rm R}(k_0\to 0,k)\right] =
- {\rm Im} \, \left[\Pi_{00}^{\rm A}(k_0\to 0,k)\right] &=&
\nn\\
&& \hspace{-4cm}
\frac{2\, g^2 \, T_F \, N_f}{\pi}\,\frac{k^0}{k}
\int_{k/2}^{+\infty}dq\,q\,
\left( \frac{n_{\rm F}(q-\mu)+n_{\rm F}(q+\mu)}{2} \right) 
\nn\\
&& \hspace{-4cm}
+ \frac{g^2 \, N_c}{\pi} \, \frac{k^0}{k}
\left[\frac{k^2}{8}\,n_{\rm B}\Big(\frac{k}{2}\Big) +
\int_{k/2}^{+\infty}dq\,q\,n_{\rm B}(q)\left( 1 - \frac{k^4}{8q^4} \right)
\right]
\,.
\label{ImPi00k0}
\eea

Taking the additional $\k\to 0$ limit here would lead to the familiar HTL self-energy. 
However, we have to keep $k$ arbitrary here since we are calculating the hard contribution. 
We can then write the contribution of our self-energy correction to the $11$ propagators as
\be
[\delta D_{00}] = \frac{ \delta D_{00}^{\rm R}(k_0,k) + \delta D_{00}^{\rm A}(k_0,k)}{2}
    + \left[\frac{1}{2} + n_{\rm B}(k_0)\right]\left(\delta D_{00}^{\rm R}(k_0,k) - \delta D_{00}^{\rm A}(k_0,k)\right),
  \label{eq:11corr}
\ee

with
$\delta D_{00}^{R/A} = -i \Pi_{00}^{R/A}(K)/{k^4}$.
 We then end up with 
 \be
\delta V^{\rm hard} = - i g^2 \, C_F \, \frac{r^2}{D-1}
\nu^{4-D} \int \frac{d^Dk}{(2\pi)^D}
\pi \delta(-k^0) \left( \frac{-i}{k^2} \right) \big[ {\cal R}_\Pi +  i \left(1 + 2n_{\rm B}(k_0)\right) {\cal I}_\Pi \big] \,.
\label{NLO1}
\ee

At this point we can integrate out the next larger scale. The leading contribution is given by an expression analogous to \Eq{eq:deltav1} in which the gluon propagators correspond to those of the HTL effective theory
\be
\delta V^{\rm soft} =- i g^2 \, C_F \, \frac{r^i}{D-1}
\nu^{4-D} \int \frac{d^Dk}{(2\pi)^D}
\frac{i}{E-h_o-k_0 +i\eta}\left[k_0^2 \,  D^{HTL}_{ii}(k_0,k) +  k^2 \,  D^{HTL}_{00}(k_0,k)
\right]r^i\,.
\label{eq:deltavbelowT}
\ee
Depending on whether the next larger scale is the Debye mass $m_D \sim g \, $max$(T,\mu)$ or the binding energy $E \sim m\als^2$ the approximations to be carried out differ. If $m_D \gg E$ then $E-h_o$ can be expanded
in the above expression. If instead $m_D \ll E$, then one can expand the self-energies in the HTL gluon propagators.
If $m_D \sim E$,
it is not possible to proceed analytically beyond extracting the UV divergences that cancel the IR ones in \Eq{VILO}, see \cite{Escobedo:2008sy,Escobedo:2010tu}. We shall not further consider this last case.

Qualitatively we can single out two cases: if $T$ is large enough we can extend the formulas obtained 
in \cite{Brambilla:2008cx,Brambilla:2010vq} for vanishing $\mu$ to the case of nonzero chemical potential, be it large $(\mu\sim T)$ or small $(\mu \ll T)$. The small $T$ case requires some extra care, as we will see in Sec. \ref{smallT}.

\subsection{Large $T$} 

We start by computing the hard contribution. 
For large $T$ ($T\gtrsim \mu$) , as long as we work at the lowest order of the expansion \Eq{octetexpand} we can use the $k^0\to 0$ limit. Then
\be 
\frac{k^0}{k}  \left[1 +2  n_{\rm B}(k_0)\right] = \frac{2 T}{k} + {\cal O}(k_0^2)\,.
\label{eq:largeTnb}
\ee

The hard contribution at finite temperature and vanishing chemical potential has been calculated in 
\cite{Brambilla:2008cx,Brambilla:2010vq}:
\begin{align}
\delta V^{\rm hard}\big\vert_{\mu=0} & = r^2 \als^2 T^3 C_F \Big\lbrace -\frac{4}{3} \zeta(3) N_c - 2 \zeta(3) N_f T_F \nn \\
              & + i \frac{2\pi}{9} \Big[ \Big( -\frac{1}{\epsilon} + \gamma + \log\pi - \log\left(\frac{T^2}{\nu^2} \right) + \frac{2}{3} - 2 \log 2 - 2 \frac{\zeta'(2)}{\zeta(2)} \Big) N_c \nn \\ 
              & \qquad +  \Big( -\frac{1}{\epsilon} + \gamma + \log\pi - \log\left(\frac{T^2}{\nu^2} \right) + \frac{2}{3} - 4 \log 2 - 2 \frac{\zeta'(2)}{\zeta(2)} \Big) N_f T_F  \Big] \Big\rbrace \nn\\
              & \equiv \delta V^R_G + \delta V^R_F + i (\delta V^I_G + \delta V^I_F) \,,
\label{eq:deltaVmuzero}
   \end{align}
 where $\zeta$ is the Riemann Zeta function and
 we can easily isolate the real and imaginary contributions $\delta V^R_G, \delta V^I_G$ coming from the gluon loops $(\sim N_c)$ from the the contributions $\delta V^R_F, \delta V^I_F$ $(\sim N_f)$ coming from the fermionic one. The former will be unchanged by the presence of a chemical potential, so we will focus on the latter.
              
\Eq{eq:deltaVmuzero} is obtained by working out first the $k$ integral using DR in \Eq{NLO1}, which helps putting all scale-less integrals to zero, then performing the $q$ integral in \Eq{RePi00k0} and \Eq{ImPi00k0}.
Note that these contributions are  suppressed by a factor $rT$ with respect to the purely real ones obtained in (\ref{deltaVlinearpluscubic}). The real part is finite, while the imaginary one has an IR log divergence.

Now let us compute these fermionic contributions at finite $\mu$. We have, 
 \begin{align}
 \delta V^R_F(T,\mu) & = -16 \pi^2 \als^2 C_F T_F \frac{N_f}{2\pi^2} \frac{r^2}{D-1}  \nu^{4-D} \int \frac{d^dk}{(2\pi)^d} \frac{1}{k^2}
 \int_0^\infty dq q \big[ \frac{n_{\rm F}(q-\mu)+n_{\rm F}(q+\mu)}{2} \big] \Big( \frac{k}{2q} - \frac{2q}{k} \Big) \log \Big\vert \frac{k - 2q}{k+2q} \Big\vert \nn\\
  & = \frac{4}{3} \als^2 C_F T_F N_f r^2 T^3 \Big[ Li\left(3, -e^{-\mu/T} \right) +  Li\left(3, -e^{\mu/T} \right) \Big] \,,  \label{eq:deltaVRF1}
 \end{align}
where $Li$ denotes the polylogarithm function.
Note also that the real part is finite. 
This will not the case for the imaginary part, which reads 
 \begin{align}
 \delta V^I_F(T,\mu) & =  -16 \pi^2 \als^2 C_F T_F  \frac{N_f}{2\pi} \frac{r^2}{D-1} \nu^{4-D}  \int \frac{d^dk}{(2\pi)^d} \frac{1}{k^2} \frac{4 T}{k} \int_{k/2}^\infty dq q
  \left( \frac{n_{\rm F}(q-\mu)+n_{\rm F}(q+\mu)}{2} \right) \nn\\
    & =  \frac{\als}{6} C_F r^2 T m_{D(F)}^2 \Bigg[ -\frac{1}{\epsilon} + \frac{2}{3} + \gamma + \log\Big(\frac{\pi}{4}\Big) - \log\Big(\frac{T^2}{\nu^2}\Big) \Bigg] \nn\\
   &  \qquad\qquad +  \frac{8\als^2}{3\pi}  C_F T_F N_f r^2 T^3 \Big[ Li^{(1,0)}(2, -e^{-\mu/T}) + Li^{(1,0)}(2, -e^{\mu/T})  \Big]\,,
    \label{eq:deltaVFmu}
 \end{align}
where we reconstructed the contribution proportional to the Debye mass, see \Eq{eq:mD}, 
 which goes together with an additional $\sim T^3$ factor 
multiplying a more involved piece containing the derivative of the polylogarithm functions with respect to their first argument.
Note that $\delta V^I$ above contributes to the decay width at leading order whereas $\delta V^R$ is subleading with respect to \Eq{deltaVlinearpluscubic}. 
Putting together the results above with the gluonic contributions in \Eq{eq:deltaVmuzero}
 and including the leading contribution \Eq{deltaVlinearpluscubic}, we can work out the corrections to the energy levels $\delta E_{nl} = \langle \delta V^R \rangle_{nl}$ 
as well as the decay rate $\Gamma_{nl} = -2  \langle \delta V^I \rangle_{nl}$ for a given $n,l$ state.
We have 
\be
\delta E_{nl}^{\rm hard} =\frac{\pi}{9}N_c C_F \als^2 T^2 \aver+\frac{2\pi}{3m}C_F\als T^2  + \frac{4}{3} \als^2 C_F  T^3 \averr \left\lbrace -\zeta (3) N_c + T_F N_f \Big[ Li\left(3, -e^{-\mu/T} \right) +  Li\left(3, -e^{\mu/T} \right) \Big]\right\rbrace\,,
\label{VRNLO}
\ee
\bea
\Gamma_{nl}^{\rm hard} &=& - \frac{1}{3} \als T C_F \averr  \left\lbrace m_{D}^2 \Big[ -\frac{1}{\epsilon} + \frac{2}{3} + \gamma + \log\Big(\frac{\pi}{4}\Big) - \log\Big(\frac{T^2}{\nu^2}\Big) \Big] \right.\nn\\
&& \left. -8 \als  T^2\left[ \frac{\pi\zeta' (2) N_c}{3\zeta (2)}- 2 \frac{ T_F N_f}{\pi}\Big( Li^{(1,0)}(2, -e^{-\mu/T}) + Li^{(1,0)}(2, -e^{\mu/T})  \Big) \right] \right\rbrace \,,
\label{VILO}
\eea
where we have introduced $\aver = a_0 [3n^2 - l(l+1)]/2$ and $\averr = a_0^2 n^2 [ 5n^2 + 1 - 3 l(l+1)]/2 $, $a_0=2/(mC_F\als)$ being the Bohr radius.

The expressions above so far hold for arbitrary $\mu$, as long as $T\gtrsim \mu$. We consider next the expansion for small $\mu$, $\mu \ll T$. 
Note that the $\mu$ dependence in $n_F(k^0)$ is analytic, and the expansion in $\mu$ does not modify its UV and IR behaviour.
Hence, the scale $\mu$ will not introduce extra singularities in our loop calculations. As a consequence, the results in this case can be obtained by just expanding in $\mu$ \Eq{VRNLO} and \Eq{VILO} above. 

For the real part, since the leading term \Eq{deltaVlinearpluscubic} does not depend on $\mu$ so it remains the same. The $\mu$ dependence arises from the next-to-leading term \Eq{eq:deltaVRF1},

 \begin{align}
 \delta V^R_F(T \gg\mu) & =
\als^2 C_F T_F N_f r^2 T^3 \Big[ -2 \zeta(3) - \frac{4}{3} \log(2) \Big(\frac{\mu}{T}\Big)^2 - \frac{1}{36}  \Big(\frac{\mu}{T}\Big)^4 + \dots  \Big] \,.
\label{RFTggmu}
 \end{align}
The first term in the expansion indeed corresponds to the $\mu=0$ result (cfr. \Eq{eq:deltaVmuzero}).
For the imaginary part, we have from \Eq{eq:deltaVFmu}
\begin{align}
\delta V^I_F(T\gg\mu)  & = -  \frac{\als}{6} C_F r^2 T m_{D(F)}^2 \Big[ \frac{1}{\epsilon} - \frac{2}{3} - \gamma - \log\Big(\frac{\nu^2}{\pi T^2}\Big)\Big] \nn\\
& - \frac{4\pi}{9} T^3 \als^2 C_f T_F N_f r^2 \Big[ 
\frac{ \zeta'(2)}{\zeta(2) } - \log\big(\frac{\pi}{4}\big) 
+ \frac{7 \zeta(3)}{8\pi^4} \Big(\frac{\mu}{T}\Big)^4 - \frac{31 \zeta(5)}{80\pi^6} \Big(\frac{\mu}{T}\Big)^6 + {\cal O}\Big(\Big(\frac{\mu}{T}\Big)^8\Big) \Big] \,,
\label{eq:deltaVFexp}
\end{align}
and
we 
recover the fermionic part  ($\sim N_f$) of \Eq{eq:deltaVmuzero} from the $\mu\to 0$ limit of \Eq{RFTggmu} and \Eq{eq:deltaVFexp}.
Keeping terms up to  ${\cal{O}}(\mu^2/T^2)$ only, we have
for the energy and the decay rate contributions
\begin{align}
\delta E_{nl}^{\rm hard} & = 
   \delta E_{nl}^{\rm hard}\Big\vert_{\mu=0} - N_f T_F C_F \frac{4\als^2 \log 2}{3} T\mu^2 \averr    \,,
\label{hRTggmu}
\end{align}
\begin{align}
\Gamma_{nl}^{\rm hard}   
 & = \Gamma_{nl}^{\rm hard}\Big\vert_{\mu=0} - \frac{4 \als^2 C_F T_F N_f T\mu^2}{3\pi} \averr  
\Big[ -\frac{1}{\epsilon} + \gamma - \log\pi - \log\left(\frac{T^2}{\nu^2} \right) + \frac{2}{3}  \Big]  \,.
\label{hITggmu}
   \end{align}
	
The expressions \Eq{VRNLO} and \Eq{VILO}, which reduce to \Eq{hRTggmu} and \Eq{hITggmu} above in the $T\gg \mu$ limit, 
are the outcome of integrating out the hard scale in the heavy quarkonium sector. In the gluonic sector the outcome is the celebrated HTL effective theory. This effective theory has exactly the same form for $\mu=0$ as for $\mu\not= 0$, the only difference being that the Debye mass acquires a $\mu$ dependence, as displayed in \Eq{eq:mD}.	
In the case $T\gg \mu$, $m_D\sim  gT$ and $m_D$ can also be expanded in a series of $(\mu/T)^2$.

Beyond the contributions 
at the hard scale, there will be additional contributions to the energy shifts and decay widths from lower scales. The form of these contributions will  depend on the relative size between $m_D$ and $m\als^2$, but not on the size of $\mu$ because of its analytic dependence. Below the hard scale $T$ we can use HTL for the light degrees of freedom.
Hence all the $\mu$ dependence will be in $m_D$  except for the case $\mu\gtrsim m_D$ in which there will be additional analytic dependences arising from the fermionic distribution function in HTL fermion loops. The latter however will be suppressed by $g^2$  factors. Let us next discuss the two most extreme cases.

\subsubsection{$m_D \gg E $ }

\label{mDcontrib1}

In this case, we can further integrate out the scale $m_D$ to get additional modifications to the potentials. These quantities basically depend on $m_D$ (except for the $T$ factor in the imaginary part coming from the $n_B$, which is unchanged), so the inclusion of a chemical potential simply amounts to considering the appropriate expression for the Debye mass in a dense medium. 
This has been worked out in \cite{Escobedo:2010tu} (see also \cite{Brambilla:2017zei}) for QED. We simply take the results from there, Eqs. (10)-(11), and 
 correct for QCD color factors. 
  We display directly the corrections to the energy shift and decay width below,
\be
\label{EgT}
{\delta E_{nl}^{(m_D)}} = C_F \frac{\als m_D^3}{6} \averr +\mathcal{O}(\als^2 r^2m_D^2T) \,,
\ee
\be
\label{DecaygT}
{ \Gamma_{nl}^{(m_D)} }=  C_F  \frac{\als T m_D^2}{3} \averr \Big(-\frac{1}{\epsilon}-\gamma+\log \pi +\log\frac{\nu^2}{m_D^2}+\frac{5}{3} \Big)+\mathcal{O}(\als^2 r^2m_D^2T) \,.
\ee

Putting together the results above with  \Eq{VRNLO}  and \Eq{VILO}, 
we get the final result for this case for the energy and the decay rate, 
\be
\delta E_{nl} =\frac{\als C_F}{3} \Big[ \frac{\pi}{3}N_c \als T^2 \aver+\frac{2\pi}{m} T^2  + 4 \als  T^3 \averr \left\lbrace -\zeta (3) N_c + T_F N_f \Big[ Li\left(3, -e^{-\mu/T} \right) +  Li\left(3, -e^{\mu/T} \right) \Big]\right\rbrace
+  \frac{ m_D^3}{2} \averr \Big] \,,
\label{RTmumDE}
\ee
\begin{align}
 \Gamma_{nl} & = - \frac{1}{3} \als T C_F \averr  \Bigg\lbrace m_{D}^2 \Big[ -1  + 2\gamma - \log4 - \log\frac{T^2}{m_D^2} \Big] \nn\\
& -8 \als  T^2\left(\frac{\pi\zeta' (2) N_c}{3\zeta (2)}- 2 \frac{ T_F N_f}{\pi}\Big[ Li^{(1,0)}(2, -e^{-\mu/T}) + Li^{(1,0)}(2, -e^{\mu/T})  \Big]\right) \Bigg\rbrace \,.
\label{ITmumDE}  
\end{align}
Note that the $1/\epsilon$ pole in the imaginary part \Eq{DecaygT} cancels with the one of \Eq{VILO}. 

For $T\gg \mu$, 
we can just  add the soft ($\sim m_D$) scale contributions, Eqs. (\ref{EgT}) and (\ref{DecaygT}),
   to the hard contribution \Eq{hRTggmu} and \Eq{hITggmu} to obtain our final result, again only up to order $\mu^2/T^2$:
\begin{align}
\delta E_{nl} & = 
 \delta E_{nl}\big\vert_{\mu=0} + \frac{\als^2}{3} C_F T_F N_f T\mu^2 \averr \Big[ -4\log 2 + \Big(\frac{3}{\pi^2} g^2 (N_c + N_f T_F) \Big)^{1/2} \Big]  \,,
\label{deltaEatmD}
\end{align}

\begin{align}
 \Gamma_{nl}  
 & = \Gamma_{nl}\Big\vert_{\mu=0}  - \frac{4 \als^2  C_F T_F N_f T\mu^2}{3\pi} \averr  
  \Big[ 2 \gamma - \log\left(\frac{T^2}{m_D^2} \right) - 1 -2 \log\pi \Big] \,.
	\label{deltaGatmD}
  \end{align}

If we consider even lower scales,  we find that the contribution at the scale $E \sim m\als^2\ll m_D$ may only be due to quasi-static magnetic photons \Eq{magnetic} and is of order $\als r^2 T E(E m_D)^{1/3}$, and hence suppressed with respect to the contributions calculated so far. Therefore, our final results in this case are \Eq{RTmumDE} and \Eq{ITmumDE}, which reduce to \Eq{deltaEatmD} and \Eq{deltaGatmD} for $T\gg\mu$.

\begin{figure}[h]{
\includegraphics[width=.4\textwidth,angle=0,scale=0.9]{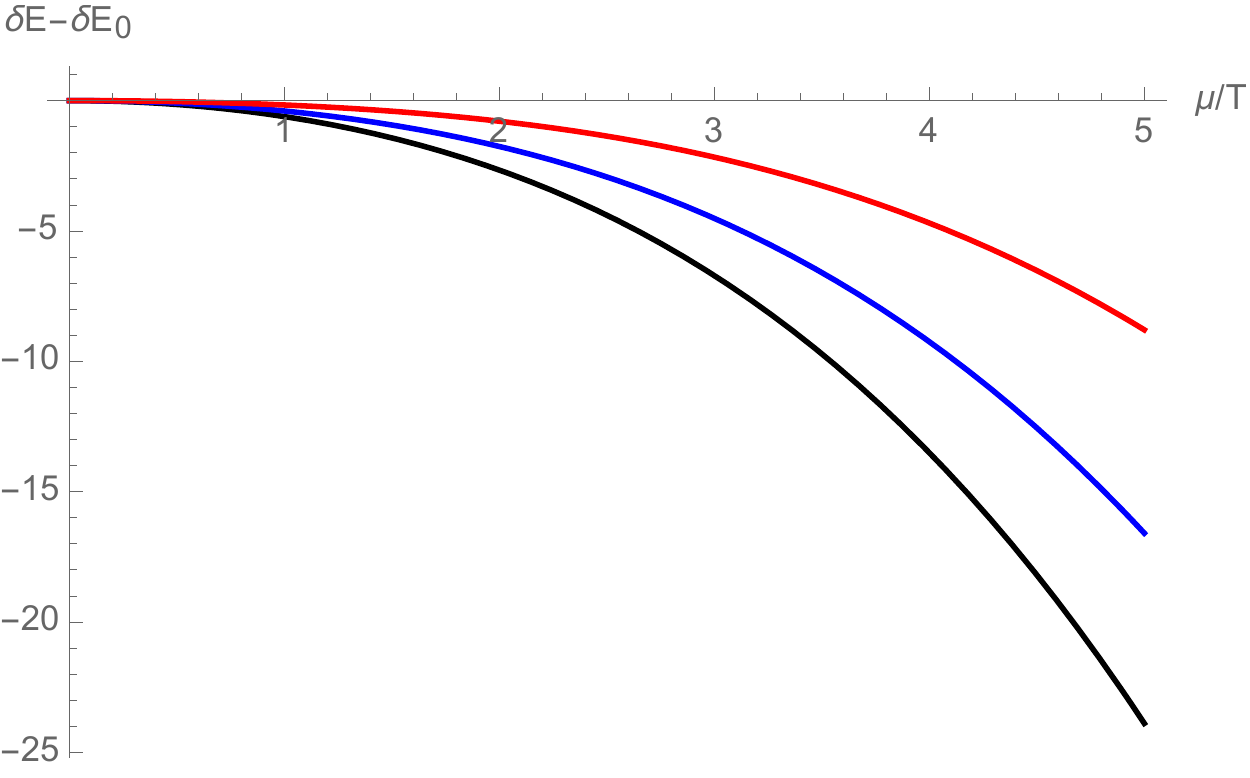}
\includegraphics[width=.4\textwidth,angle=0,scale=0.9]{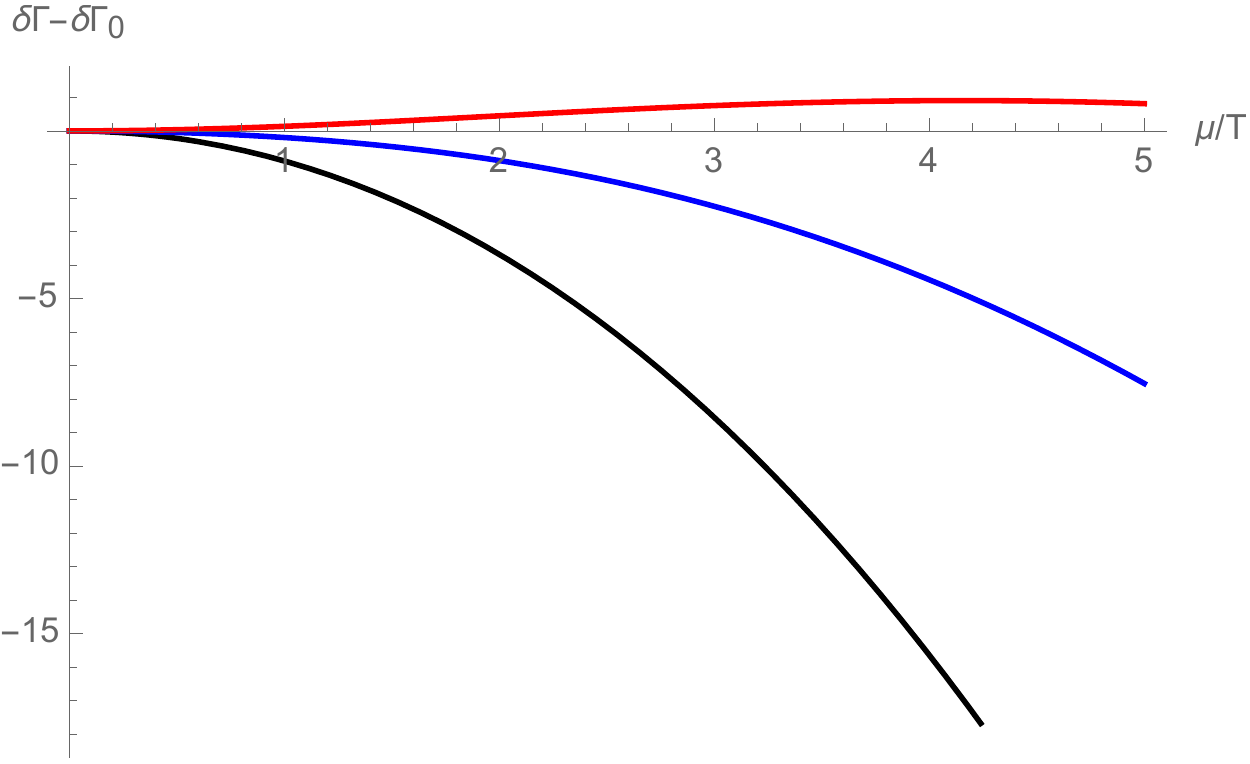}
\caption{ Quarkonium energy shift and decay rate for the ground state including $m_D$ scale corrections as function of the ratio $\mu/T$, for $N_f=2$ and different values of $\als$. More specifically, we plot in the left panel the result  $(\delta E_{10} - \delta E_{10}\vert_{\mu=0}) / (4\als^2 C_F T^3 \averrz/3)$, and in the right one  $(\Gamma_{10} - \Gamma_{10}\vert_{\mu=0}) / (-4\pi \als^2 T^3 C_F \averrz/9 )$.
 Curves are for $\als$=0.01 (black), 0.1 (blue) and 0.3 (red). 
 \label{fig:mDcontribs}}}
\end{figure}

In order to get a feeling on the contributions computed in this section, we plot in Fig. \ref{fig:mDcontribs} the results for the energy shift and the decay rate as function of the chemical potential for different values of $\als$. 

\subsubsection{$E \gg m_D $ }

\label{a2}

If $E\sim m\als^2 > m_D$ we should be integrating out this scale first rather than the Debye mass. This has been worked out at $\mu=0$ in \cite{Escobedo:2010tu} (Eqs. (6)-(7)) for QED, and in
\cite{Brambilla:2010vq} for QCD. 
In this case the denominator \Eq{octetexpand} cannot be expanded, so that we are no longer fixed to the $k^0 \to 0$ limit, but 
we can still make use of the expansion \Eq{eq:largeTnb} for the bosonic occupation number.
Furthermore, we can employ the HTL gluon self-energies expanded in powers of $m_D/E$.
 The leading energy shift is given by the longitudinal gluon contribution, Eq (5.18) in 
\cite{Brambilla:2010vq}, 
\be
\delta E_{nl}^{ (E)}=-\frac{\pi\als C_F\ Tm_D^2}{3} \averr\,,
\label{RTmuEmDE}
\ee
whereas both longitudinal and transverse gluons contribute to the decay width, which is given by (5.25) in \cite{Brambilla:2010vq},
\begin{align}
\Gamma_{nl}^{(E)}&=
C^{(1)}_{nl} T  + C^{(2)}_{nl}   -\frac{\als\cf Tm_D^2}{3}\left(\frac{1}{\epsilon}
+\log\frac{E_1^2}{\nu^2}+\gamma-\frac{11}{3}-\log\pi+\log4\right) \averr \nn\\
 &+\frac{2\als\cf Tm_D^2}{3}\frac{C_F^2\als^2}{E_n^2}\,I_{n,l}  \label{ITmuEmDE}\;,
\end{align}
where $E_n=-m\cf^2\als^2/(4n^2)$, $n=1,2,\dots$, is the energy of the state and
$I_{n,l}$ a numerical constant dependent on the $n,l$ state given in \cite{Brambilla:2010vq}. We introduced the shorthand notation 
\begin{align}
C_{nl}^{(1)} & = \frac{1}{3}N_c^2C_F\als^3-\frac{16}{3m}C_F\als E_n+\frac{8}{3}N_cC_F\als^2\frac{1}{mn^2a_0} \qquad {\rm and}  \\
C_{nl}^{(2)} & = \frac{2E_n\als^3}{3}\left\{\frac{4 C_F^3\delta_{l0} }{n}+N_c \cf^2
\left(\frac{8}{n (2l+1)}-\frac{1}{n^2} - \frac{2\delta_{l0}}{ n }   \right)
+\frac{2N_c^2C_F}{n(2l+1)}+\frac{N_c^3}{4}\right\}
 \,,
\end{align}
the latter being a subleading $T$-independent contribution which we will neglect in the following.

Again the generalization to finite $\mu$ is straightforward: we can clearly distinguish single factors of $T$, which come from the expansion of the bosonic distribution function in the gluon propagator and thus are unchanged by the introduction of density,
whereas all the remaining medium dependence is expressed in terms of the Debye mass. We can just replace $m_D$ by the appropriate $\mu-$dependent value for the Debye mass,  given by \Eq{eq:mD} .

\begin{figure}[h]{
\includegraphics[width=.4\textwidth,angle=0,scale=0.9]{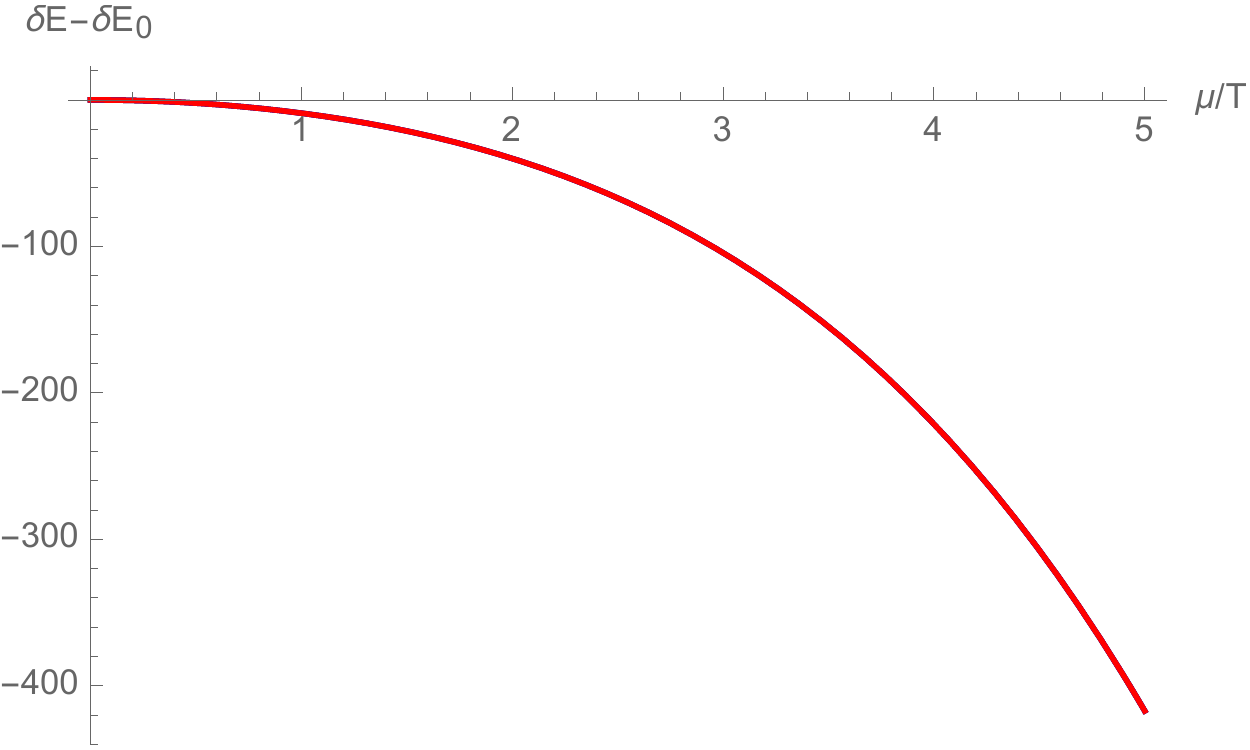}
\includegraphics[width=.4\textwidth,angle=0,scale=0.9]{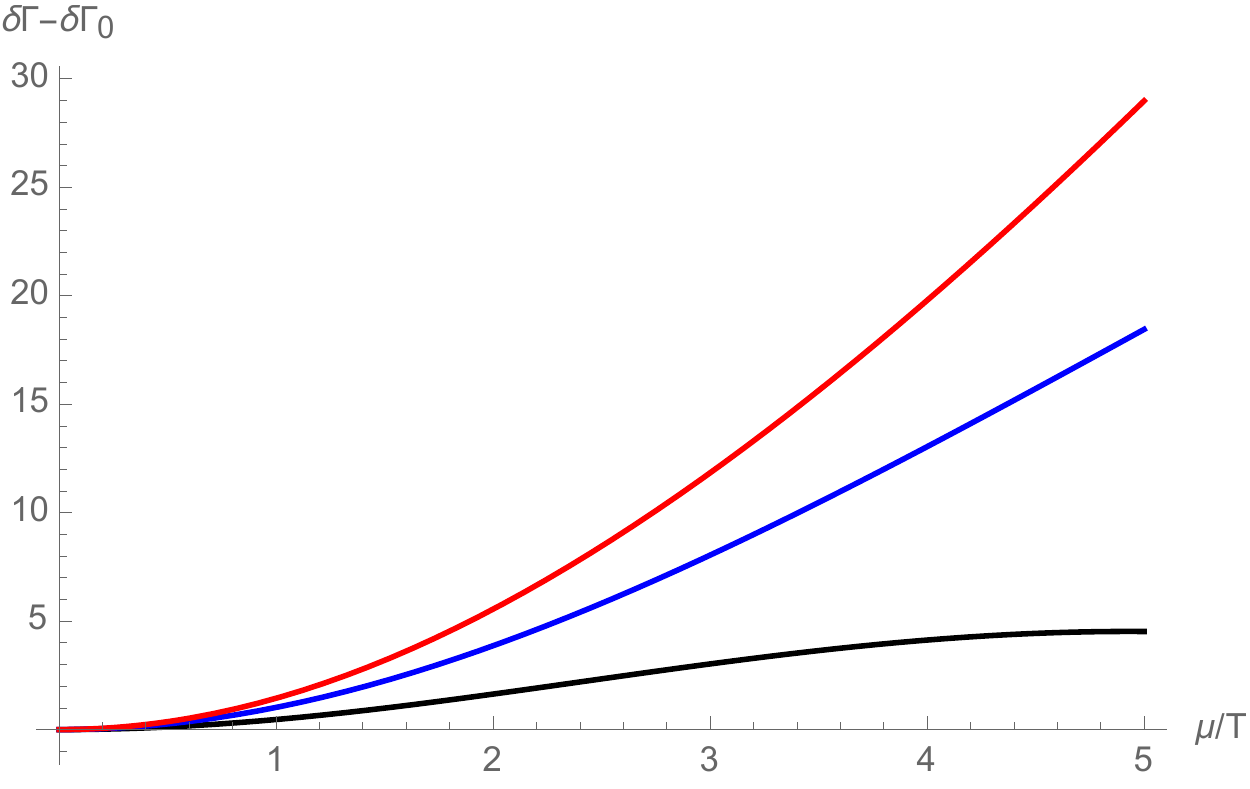}  
\caption{ Quarkonium energy shift and decay rate for the ground state as function of the ratio $\mu/T$ including $E$ scale corrections, for $N_f=2$. We plot in the left panel the result  $(\delta E_{10} - \delta E_{10}\vert_{\mu=0}) / (4\als^2 C_F T^3 \averrz/3)$, and in the right one  $(\Gamma_{10} - \Gamma_{10}\vert_{\mu=0}) / (-4\pi \als^2 T^3 C_F \averrz/9 )$. This time our ratio for the energy shift turns out to be $\als$-independent, while for the decay rate 
 we need to specify the value of the ratio $T/m\als^2$: here we chose for illustration 2 (black), 5 (blue) and 10 (red).
 \label{fig:Econtribs}}}
\end{figure}

We then get out final result for this case by adding the above expressions to \Eq{VRNLO} and \Eq{VILO}, 
\begin{align}
\delta E_{nl} & = 
\frac{\pi}{9}N_c C_F \als^2 T^2 \aver+\frac{2\pi}{3m}C_F\als T^2  \nn\\ 
&+ \frac{4}{3} \als^2 C_F  T^3 \averr \left\lbrace -\zeta (3) N_c + T_F N_f \Big[ Li\left(3, -e^{-\mu/T} \right) +  Li\left(3, -e^{\mu/T} \right) \Big]\right\rbrace 
-\frac{\pi\als C_F\ Tm_D^2}{3} \averr \,,
\label{RTmuEmD}
\end{align}
\begin{align}
\Gamma_{nl}&=
- \frac{1}{3} \als T C_F \averr  \left\lbrace m_{D}^2 \Big[ -3 + 2\gamma - \log\frac{T^2}{E_1^2} \Big] \right.\nn\\
& \left.- 8 \als T^2\left(\frac{\pi\zeta' (2) N_c}{3\zeta (2)}- 2\frac{ T_F N_f}{\pi}\Big[ Li^{(1,0)}(2, -e^{-\mu/T}) + Li^{(1,0)}(2, -e^{\mu/T})  \Big]\right)\right\rbrace 
 + C_{nl}^{(1)} T 
 +\frac{2\als^3\cf^3 Tm_D^2}{3E_n^2}\,I_{n,l}\;.
\label{ITmuEmD}
\end{align}
Note that the $1/\epsilon$ poles in the imaginary part \Eq{DecaygT} cancels with the one of \Eq{ITmuEmDE}. For $T\gg \mu$,
we can just  add \Eq{RTmuEmDE} and \Eq{ITmuEmDE} 
to the hard contribution \Eq{hRTggmu} and \Eq{hITggmu} to obtain our final result, again only keeping the leading correction in $\mu/T$: 
\begin{align}
\delta E_{nl} & = 
  \delta E_{nl}\Big\vert_{\mu=0} - \frac{4}{3} C_F T_F N_f \als^2 T\mu^2 \averr \Big[ \log 2 + 1 \Big] \,,
\label{RTmuEmDexp}
\end{align}

\begin{align}
\Gamma_{nl}&=
\Gamma_{nl}\Big\vert_{\mu=0} - \frac{4\als^2 C_F T_F N_f T\mu^2}{3 \pi}  \Big[  \averr \Big(2\gamma + \log 4 - \log\frac{T^2}{E_1^2} - 3  -2 \log\pi \Big) - 2 \frac{C_F^2 \als^2}{E_n^2} I_{n,l} \Big] \,.
\label{ITmuEmDexp}
\end{align}
If we consider lower scales, in this case the scale $m_D$, we find that it
 gives contributions of the order $\als r^2 T m_D^3/E$ which are suppressed with respect to the ones calculated so far. Hence, our final results in this case are given by \Eq{RTmuEmD} and \Eq{ITmuEmD}, which reduce to \Eq{RTmuEmDexp} and \Eq{ITmuEmDexp} for $T\gg \mu$.
We plot the resulting expressions as function of the ratio $\mu/T$ in Fig.\ref{fig:Econtribs}.

\subsection{Small $T$}
\label{smallT}

Let us next consider the case $\mu \gg T$.
This case is technically more involved as it cannot be obtained by just taking the small $T$ limit of the 
{ general results in Sec. \ref{hard}} (recall that the distribution functions are not analytic in $T$). 
Below the scale $\mu$, HTL must be used but the approximation $N_B(k_0)\sim 2T/k_0$ for the Bose distribution function does not hold in general. This is because the constraint $k_0 \ll \mu$ still allows $k_0 \gtrsim T$. Let us study the two extreme cases, $T\gg E$ and $T\ll E$ below.

\subsubsection{$T \gg E$ }
 
{ Our starting point here is still formulas (\ref{NLO1}) and (\ref{eq:deltavbelowT}) for the hard and soft contributions respectively}.
The contributions at the hard ($\mu$) scale are now restricted to the quark loop in \Fig{fig:self2}, 
see sec. \ref{2lh}. The leading contribution 
can be worked out 
by just replacing in  \Eq{eq:deltaVRF1} and \Eq{eq:deltaVFmu}
$  n_F(q - \mu) + n_F(q + \mu) \rightarrow \theta(\mu - q)$. The leading $T$ dependence requires more effort, one can nevertheless  work it out. 
 The real part becomes
\be
\delta V^R_F(T\ll\mu) = -\frac{2}{9} \als^2 C_F T_F N_f r^2 \left(\mu^3 +\pi^2 T^2\mu\right)\,,
\label{LORemu}
\ee
up to exponentially suppressed terms,  and 
the imaginary part reads 
\be
\delta V^I_F(T\ll\mu) =  - \frac{2 \als^2 C_F T_F {N_f} {r^2} T 
}{3 \pi} \Big[ \left(\mu^2+\frac{\pi^2 T^2}{3}\right)  \left(\frac{1}{ \epsilon}+ \log \frac{\mu^2}{\nu^2} + \gamma -\log\frac{\pi}{4} - \frac{11}{3} \right) - \pi^2 T^2  +{\cal O}(T^4/\mu^2) \Big]\,.
\label{hardBE}
\ee
We see that the overall factor of $T$ coming from the bosonic occupation number makes this imaginary contribution parametrically smaller than its real counterpart. 
At scales below $\mu$ the HTL effective theory must be used for gluons and light quarks. 
If $ T\gg m_D \sim g\mu \,, E$, we can next integrate $T$ out.
The selfenergies can then be expanded in powers of $m_D$ in the HTL propagators, which at leading order {reduce to the bare ones}. Hence, we obtain an extra contribution, which coincides with \Eq{deltaVlinearpluscubic}. Since the last result is finite, we still need to integrate out the next larger scale in order to cancel the $1/\epsilon$ pole in \Eq{hardBE}.
This is done
in Sec. \ref{TmDE} and \ref{TEmD} below. We address the case $m_D \gg T\gg E$ in Sec. \ref{mDTE}.
 Note that for $\mu \gg T$ the gluon distribution functions are exponentially suppressed at the hard scale, and hence they do not contribute to the HTL selfenergies. Therefore, in the following subsections, all the Debye masses will only have the fermionic contributions. 
\vspace{0.5cm}
\paragraph{The case $T\gg m_D\sim g\mu \gg E$}\label{TmDE} $\quad$

\vspace{0.5cm}

We can take the result of integrating out $m_D$ from \Eq{EgT} and \Eq{DecaygT}. Putting everything together we finally obtain 
\begin{align}
\delta E_{nl} &= \frac{\pi}{9}N_c C_F \als^2 T^2 \aver+\frac{2\pi}{3m}C_F\als T^2  -\frac{2}{9} \als^2 C_F T_F N_f \averr \left(\mu^3 +\pi^2 T^2\mu\right) \nn\\
& + C_F \frac{\als m_{D(F)}^3}{6} \averr \,,
\end{align}

\begin{align}
\Gamma_{nl} 
 & = \frac{1}{3} \als T C_F\averr  \Big[m_{D(F)}^2 \Big( \log 4 - 2 + \log \frac{\mu^2}{m_{D(F)}^2}  \Big) - 4\pi\als  T_F {N_f}  T^2 \Big]  \,.
\end{align}
 Note that, parametrically, in the energy shift, the two first terms and the third term in the first line compete to be the leading contribution, whereas the remaining ones are suppressed. The last term in the decay width is also suppressed.

\vspace{0.5cm}
\paragraph{The case $T\gg E\gg m_D\sim g\mu$}\label{TEmD} $\quad$

\vspace{0.5cm}

We can take the result of integrating out $E$ from \Eq{RTmuEmDE} and \Eq{ITmuEmDE}. Putting everything together we finally obtain 
\begin{align}
\delta E_{nl} &= \frac{\pi}{9}N_c C_F \als^2 T^2 \aver+\frac{2\pi}{3m}C_F\als T^2  -\frac{2}{9} \als^2 C_F T_F N_f \averr \left(\mu^3 +\pi^2 T^2\mu\right) \nn\\
&  -\frac{\pi\als C_F\ Tm_{D(F)}^2}{3} \averr \,,
\end{align}

\begin{align}
\Gamma_{nl} 
 &= \frac{\als\cf T}{3} \averr  \Big[m_{D(F)}^2 \log \frac{\mu^2}{E_1^2} - 4\pi\als  T_F {N_f}  T^2 \Big] 
  +
 C^{(1)}_{nl} T   +\frac{2\als\cf Tm_{D(F)}^2}{3}\frac{C_F^2\als^2}{E_n^2}\,I_{n,l} \,.
\end{align}

Parametrically, in the energy shift, all terms may compete for the leading order, except for term linear in $\mu$, which is always smaller than the term cubic in $\mu$. In the decay width, the terms proportional to $C^{(1)}_{nl}$ and $m_{D(F)}^2$ are the leading and next-to-leading ones respectively, while the one proportional to $T^3$ is suppressed.

\vspace{0.5cm}
\paragraph{
 The case $\mu \gg m_D \sim g\mu \gg T\gg E$ \label{mDTE}} $\quad$

\vspace{0.5cm}

\begin{figure}{
\includegraphics[width=.4\textwidth,angle=0,scale=0.5]{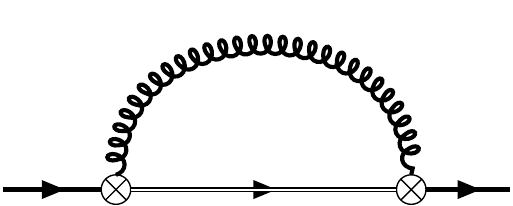}
\caption{  
The thick curly line denotes the low energy magnetic gluon propagator in \Eq{magnetic}. The solid and double lines are the suitable color singlet and color octet quarkonium propagators, and the crossed dots 
chromoelectric dipole vertices. \label{mg}}}
\end{figure}

In this case, the next scale to be integrated out is $m_D$.
This produces the
same contribution as in the general case, namely \Eq{EgT} and \Eq{DecaygT} with $m_{D(F)}$ instead of the full $m_D$, as in the previous subsection.
 Putting everything together we obtain 
\begin{align}
\delta E_{nl} &= -\frac{2}{9} \als^2 C_F T_F N_f \averr \left(\mu^3 +\pi^2 T^2\mu\right) 
+ C_F \frac{\als m_{D(F)}^3}{6} \averr \,,
\label{deltaEmDTE}
\end{align}

\begin{align}
\Gamma_{nl}
 & = \frac{1}{3} \als C_F\averr  \Bigg\lbrace T m_{D(F)}^2 \Big[ \log 4 - 2 + \log\Big(\frac{\mu^2}{m_{D(F)}^2} \Big) \Big] - 4\pi\als  T_F {N_f}  T^3 \Bigg\rbrace  \,.
\end{align}
 Parametrically, the first term both in the energy shift and in the decay width above is the leading one. Concerning the $T$-dependent terms, since we have not considered contributions at the scale $T$ so far, we may wonder whether the terms above are the most important ones, or contributions at lower scales may provide larger $T$-dependent terms. In order to resolve this question,
let us next consider the contributions at lower scales.
 Only quasi-static magnetic modes survive below $m_D$. These are obtained by approximating the transverse HTL self-energy to the case $k^0\ll k\ll m_D$, see \Eq{magnetic}.
When $m_D\gg T$, these modes contribute at lower scales from the diagram in \Fig{mg}.
 At the scale $T$, their contribution is of the order $\als r^2 T^2(T m_D^2)^{1/3}$, and hence it becomes the most important $T$-dependent contribution to the real part of the potential \footnote{There might be competing logarithmic $T$-dependent contributions of order $\als^2 r^2 E \mu^2$ from the region $k_0\sim T$, $k\sim \mu$, similar to those displayed in the Appendix \ref{loglog}.}. It reads
\be
\delta V = -g^2 C_F \frac{2}{3}  (-3) \Big[ \frac{1}{m} + \frac{1}{6} N_c \als r \Big] \frac{T}{6\pi^2} \Big( \frac{\pi m_{D(F)}^2 T}{4} \Big)^{1/3} \frac{\Gamma(4/3)\zeta(4/3)}{\cos(\pi/6)} \,,
\ee
which produces a further energy shift to be added to \Eq{deltaEmDTE}, 
\be
\delta E_{nl} =   \frac{4C_F\als}{3\pi}  \Big[ \frac{1}{m} + \frac{1}{6} N_c \als \aver \Big] T \Big( \frac{\pi m_{D(F)}^2 T}{4} \Big)^{1/3} \frac{\Gamma(4/3)\zeta(4/3)}{\cos(\pi/6)}\,.
\ee
There are also contributions at the scale $E$ from the quasi static magnetic modes, which are of order 
$\als r^2 TE (Em_D^2)^{1/3}$, and hence suppressed with respect to the ones considered so far.

\subsubsection{$E \gg T$}
\label{sec:EgtrTlow}

The case $T \ll E\sim m\als^2$ deserves a special treatment.
 Let us focus first on the 
the longitudinal contribution of \Eq{eq:deltav1}, which becomes  
\begin{align}
\delta V  
 &= - i g^2 \, C_F \, \frac{r^i}{D-1}
\nu^{4-D} \int \frac{d^Dk}{(2\pi)^D} \frac{i}{a -k_0+i\eta} \frac{k^2}{2} \Big[ D_{00}^{\rm R}(k_0,k) +D_{00}^{\rm A}(k_0,k)  +  N_B(k^0) (D_{00}^{\rm R}(k_0,k) - D_{00}^{\rm A}(k_0,k)) \Big]r^i 
\,,
\label{eq:deltavLowT1}
\end{align}
where we introduced for brevity $a \equiv E- h_0 $ 
{and we are not assuming any specific form for the longitudinal gluon propagator yet}.  
In order to single out  real and imaginary parts we consider the combinations $\delta V^{R} = (\delta V + \delta V^*)/2$ and $\delta V^I = -i (\delta V - \delta V^*)/2$.

 After writing the denominator as $\frac{i}{a -k_0+i\eta} = -i {\cal {P}}\frac{1}{k_0-a} + \pi \delta(a-k^0)$, we thus get to
 \begin{align}\label{deltaVR}
 \delta V^R & = -i g^2 C_F \frac{r^i}{D-1}
\nu^{4-D}\int \frac{d^dk}{(2\pi)^d} k^2 \Big\lbrace \frac{1}{2} \Big[ D_{00}^R(a,k) + D_{00}^A(a,k) \Big] + i {\cal P} \int \frac{dk^0}{2\pi} \Big( \frac{1}{a-k^0} \Big) N_B(k^0)  \Big[ D_{00}^R(k^0,k) - D_{00}^A(k^0,k) \Big] \Big\rbrace r^i  \,, \\
\delta V^I & = -i g^2 C_F \frac{r^i}{D-1}
\nu^{4-D} \int \frac{d^dk}{(2\pi)^d} k^2 \Big\lbrace  i {\cal P} \int \frac{dk^0}{2\pi} \Big(\frac{1}{a-k^0} \Big) \Big[ D_{00}^R(k_0,k) + D_{00}^A(k_0,k) \Big] + \frac{1}{2} N_B(a)  \Big[ D_{00}^R(a,k) - D_{00}^A(a,k)\Big] \Big\rbrace r^i \,.
\end{align}
{The $k_0$ integral of $\delta V^I$ can be carried out if we assume that all the singularities (poles or branch points) in $D_{00}(k_0,k)$ are on the real axis \cite{Bellac:2011kqa} (this can be explicitely verified for the approximations we use, namely for the one loop self-energy and for the HTL propagator). The integral is done by writing the principal value as ${\cal P}/(a-k_0)=(1/(a-k_0+i\eta)+1/(a-k_0-i\eta))/2$. Then we get four terms. Two of them have all the singularities in the same complex half plane and hence vanish. In the remaining two terms the singularity of the $1/(a-k_0\pm i\eta)$ is in the opposite half plane as the ones of the accompanying propagator. Hence, by closing the path so that only the singularities in $1/(a-k_0\pm i\eta)$ are enclosed we obtain}
\begin{align}
\delta V^I   = - i g^2 C_F  \frac{r^i}{D-1}
\nu^{4-D} \int \frac{d^dk}{(2\pi)^d} \frac{k^2}{4} \Big[ D_{00}^R(a,k) -  D_{00}^A(a,k)  +  N_B(a)\left(D_{00}^R(a,k)  -  D_{00}^A(a,k) \right)  \Big]r^i %= 0
\,.
\label{eq:deltavLowT2}
\end{align}
{The expressions used in the previous sections can also be obtained from \Eq{deltaVR} and \Eq{eq:deltavLowT2}. For instance, \Eq{NLO1} corresponds to substituting the gluon propagators by the contribution of the one-loop self-energy to them. In the hard contribution, $a$ is small and can be set to zero at LO, which is reminiscent of the Dirac delta in \Eq{NLO1} (recall that $N_B(a)\sim 2T/a$, so there is also a contribution from the imaginary part). The longitudinal part of \Eq{eq:deltavbelowT} is obtained by replacing the propagators above by their HTL expressions.}

{Coming back to the $E\gg T$ case, recall} that $a= E-h_o \sim E \gg T$. We can then approximate $N_B(a)\sim$ sgn$(a)=-1$, for any $a<0$, as it is the case for a bound state, since $E<0$ and $h_o$ is positive definite. 
Hence the imaginary part is zero {($\delta V^I=0$)}, irrespectively of the form of the longitudinal gluon propagator.

A similar reasoning can be done with the transverse contribution, which also leads to a vanishing contribution for the imaginary part. The real part can be obtained from \Eq{deltaVR} by replacing $k^2\to k_0^2$ and $D_{00}^{R,A}(k_0,k)\to D_{ii}^{R,A}(k_0,k)$.  
Therefore no imaginary part to the potential is generated when $T$ is smaller than the binding energy scale.  
Note that the same argument is valid both for the two-loop hard contribution as well as for the HTL one, as it does not depend on the the details of the gluon propagator.

Let us first consider the hard contribution $k\sim \mu$ to the real part. The first term in \Eq{deltaVR} gives \Eq{LORemu}, as expected, since $a\ll \mu$. The second term is subleading. Indeed, for $k_0\sim k \sim \mu$, the denominator can be expanded in $a$, the would-be-leading order vanishes (odd in $k_0$), and hence the leading contribution is $a/\mu$ suppressed. For $k_0 \ll k\sim \mu$, the difference between retarded and advanced propagators is proportional to $k_0/k$, and hence also suppressed. Nevertheless, the region $k_0 \sim T$ provides the leading $T$-dependent contribution,
\be
\delta V^R_{\rm hard}\vert_{T-{\rm dependent}}= \frac{\pi\als C_F m_{D(F)}^2 T^2}{9}\left(\frac{1}{\epsilon} -\log \pi-\psi\left(\frac{3}{2}\right)-\frac{7}{6} +2\log \frac{\mu}{\nu} \right) r^i\frac{1}{E-h_o}r^i\,,
\label{deltaVRhard}
\ee
where $\psi (z)$ is the Digamma function.
Note that $\delta V^R_{\rm hard}\vert_{T-{\rm dependent}}$ is not really a potential, as it depends on the external energy $E$.
The soft regions, $k_0, k\ll \mu$, give subleading contributions, but they do contribute to the leading $T$-dependence. Let us display the two extreme cases,
\vspace{0.5cm}
\paragraph{The case $\mu\gg m_D\,, E\gg T$} $\quad$ \vspace{0.5cm}

 Let us next consider the contributions at the scale $k\sim m_D$. The first term in \Eq{deltaVR} gives \Eq{EgT}, which together with \Eq{hardBE} leads to \Eq{deltaEmDTE}. However, the leading $T$-dependence is not given by this expression but by the region $k\sim m_D$, $k_0\sim T$ in the second term of \Eq{deltaVR},

\be
\delta V^R_{\rm soft}\vert_{T-{\rm dependent}}=\frac{\pi\als C_F m_{D(F)}^2 T^2}{9}\left(-\frac{1}{\epsilon} +\log \pi+\psi\left(\frac{3}{2}\right)-\frac{1}{3} -2\log \frac{m_{D(F)}}{\nu} \right) r^i\frac{1}{E-h_o}r^i\,,
\ee
which together with \Eq{deltaVRhard} leads to the following $T$-dependent energy shift,

\be
\label{pmuETTdep}
\delta E_{nl}\vert_{T-{\rm dependent}}=-\frac{\pi\als C_F m_{D(F)}^2 T^2}{9}\left(\frac{3}{2} +2\log \frac{m_{D(F)}}{\mu} \right) 
\Big\langle r^i\frac{1}{E-h_o}r^i\Big\rangle_{nl} \,.
\ee
The matrix element above has been calculated in \cite{Voloshin:1979uv,Leutwyler:1980tn} (see also \cite{Pineda:1996uk}). The contribution of the transverse photons is $T^2/m_D^2$ suppressed with respect the one of the longitudinal gluons that we have just displayed.

\vspace{0.5cm}
\paragraph{The case $\mu\gg E\gg T\gg m_D$} $\quad$ \vspace{0.5cm}

In this case  the region $k\sim k_0\sim T$ gives the leading $T$-dependence. In fact, it is due to the transverse gluons because their propagator at tree level already contributes. It gives the same result as for the $\mu=0$ case, namely,

\be
\delta V^R_{\rm soft}\vert_{T-{\rm dependent}}=\frac{g^2C_F\pi^2 T^4}{45}r^i\frac{1}{E-h_o}r^i\,.
\label{Tonly}
\ee
The abelian limit of the expression above agrees with the one of \cite{Escobedo:2008sy}. The size of this term is parametrically larger than the hard contribution in \Eq{deltaVRhard}. 
Nevertheless, it is important to carry out the calculation at the soft scale that cancels the $1/\epsilon$ pole in \Eq{deltaVRhard}. This is realized by the longitudinal gluons at the scale $k_0\sim k\sim T$, which give 
\be
\delta V^R_{\rm soft}\vert_{T-{\rm dependent, subleading}}= \frac{\pi\als C_F m_{D(F)}^2 T^2}{9}\left(-\frac{1}{\epsilon} +\log \pi+\psi\left(\frac{3}{2}\right)-\frac{4}{3} +2\gamma
- 2 \frac{\zeta'(2)}{\zeta(2)}
-2\log \frac{T}{2\nu} \right) r^i\frac{1}{E-h_o}r^i\,.
\label{deltaVRsoft2}
\ee
Putting together \Eq{Tonly}, \Eq{deltaVRhard} and \Eq{deltaVRsoft2},
 we obtain for the $T$-dependent energy shift in this case, 
\be
\delta E_{nl}\vert_{T-{\rm dependent}}=\left[\frac{g^2C_F\pi^2 T^4}{45}+\frac{\pi\als C_F m_{D(F)}^2 T^2}{9}\left(-\frac{5}{2} +2\gamma
- 2 \frac{\zeta'(2)}{\zeta(2)}
 -2\log \frac{T}{2\mu} \right)\right]
\Big\langle r^i\frac{1}{E-h_o}r^i\Big\rangle_{nl} \,.
\ee

\section{The case  $m \gg {\rm max}(T,\mu) \gg p \sim m_D \gg E$} 
\label{TmupE}

So far the energy scales associated with the thermal medium were assumed to be smaller than the typical momentum exchanges $p$ between the constituents of the bound state. This implies that thermal effects can be treated as perturbations to the bound state dynamics. The melting of the bound state may still occur, because it can develop a medium decay width comparable to the binding energy. One may wonder however, in which conditions the medium effects will be so strong that they will affect the leading-order bound state dynamics, namely the leading order potential. When $p\sim$ max$(T,\mu)$, this is not the case yet. This is because the longitudinal gluon propagator is not sensitive to the medium at tree level, and hence the Coulomb-like potential remains as the LO potential. The one loop correction is
suppressed by a $g^2$ factor, and hence medium effects are still a perturbation.

For $\mu=0$, this case is analyzed
in Sec. IV of \cite{Escobedo:2008sy} and 
in Sec. IIb/Appendix D of \cite{Escobedo:2010tu} for QED.
We shall not develop it further, since it does not bring in any qualitative difference with respect to the previous section. In contrast, the case max$(T,\mu)\gg p\sim m_D$ introduces modifications in the LO potential, and hence in the full bound state dynamics. For $\mu=0$, in the static limit of QCD ($m\to \infty$, $p\equiv 1/r$), this case was addressed in \cite{Laine:2006ns} and \cite{Brambilla:2008cx},
and in the full dynamical case of QED in \cite{Escobedo:2008sy,Escobedo:2010tu}. In the following we extend these results to finite chemical potential.

The suitable starting point now is Non-Relativistic QCD (NRQCD)\cite{Caswell:1985ui,Bodwin:1994jh}, since the heavy quark mass is still larger than the remaining scales in the problem, and hence it can be integrated out. We will only need the leading order Lagrangian,
\begin{eqnarray}
{\cal L}_{\textrm{pNRQCD}} &=
&
- \frac{1}{4} F^a_{\mu \nu} F^{a\,\mu \nu}
+ \sum_{i=1}^{N_f}\bar{q}_i\,iD\!\!\!\!/\,q_i +\left(\psi^\dagger\left(iD_0+\frac{\mathbf{D}^2}{2m}+\cdots \right)\psi+{\rm c.c.}\right)\,,
\label{NRQCD}
\eea
where $\psi$ is a non-relativistic field that annihilates heavy quarks, and c.c. stands for the charge conjugated term, namely the analogous terms for the heavy antiquarks, see \cite{Brambilla:2004jw,Pineda:2011dg}.

\subsection{Integrating out the hard scale} 

In the gluon and light quark sector, the integration of the largest scale max$(T,\mu)$ produces the HTL effective theory. In the heavy quark sector, it produces a shift of the heavy quark mass $\delta m$.
In the static limit, 
 the leading contribution corresponds to the two-loop diagram in \Fig{2lm}, which is ${\cal O}(\als^2$max$(T,\mu))$ and  turns out to be suppressed by a factor of $g$ with respect to lower energy contributions.
However, when $1/m$ corrections are considered, there is a leading order contribution from the diagram of Fig. \ref{1lmm} provided that $T$ is the largest scale, $\delta m \sim \als T^2/m\sim m \als^2\sim E$ \cite{Escobedo:2008sy},
\be
\delta m=\frac{\pi C_F \als T^2}{3m} \,.
\label{dm}
\ee

\begin{figure}{
\includegraphics[width=.4\textwidth,angle=0,scale=0.5]{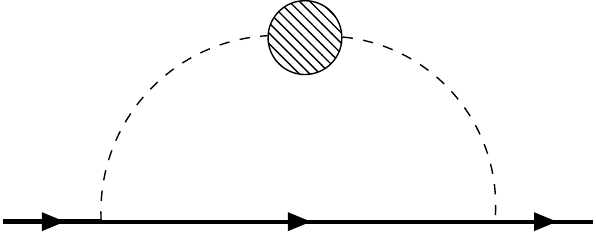}
\caption{Leading contribution to the heavy quark self-energy at the hard scale in the static limit. The solid and dashed lines denote the heavy quark and the longitudinal gluon propagators respectively, and the blob the longitudinal gluon selfenergy.} 
\label{2lm}}
\end{figure}

\begin{figure}{
\includegraphics[width=.4\textwidth,angle=0,scale=0.5]{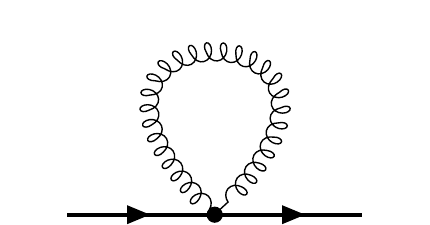}
\caption{
 \label{1lmm} Tadpole contribution to the heavy quark self-energy from the $\mathbf{D}^2/2m$ term in \Eq{NRQCD}. The solid and curly lines denote the heavy quark and transverse gluon propagators respectively. 
 }}
\end{figure}

We now proceed to integrating out the lower scales. As before, it is useful to treat separately the cases in which $T$ is large and small, respectively.

\subsection{Large $T$ ($T\gtrsim \mu$)}

This corresponds to calculating the mass shift  and potentials (${\cal O}(\als m_D \sim m\als^2\sim E$) ) using HTL. The result can then be just read from \cite{Laine:2006ns,Brambilla:2008cx}.
For the mass shift we get, 
\be
\delta m = -\frac{C_F \als}{2} \left( m_D + i T \right)  \,,
\label{dmhtl}
\ee
while for the potential shift, 
\begin{align}
V_s(r)
= -C_F\,\frac{\als}{r}\,e^{-m_Dr}
+ iC_F\,\als\, T\,\frac{2}{rm_D}\int_0^\infty dx \,\frac{\sin(m_Dr\,x)}{(x^2+1)^2} \,,
\label{V}
\end{align}
where now $m_D$ may depend on both $\mu$ and $T$. Recall that the potential develops an imaginary part, first uncovered in \cite{Laine:2006ns}. If $T\sim \mu$, our final result for the potential plus mass shift is just the addition of twice the (complex) mass shifts in \Eq{dm} and in \Eq{dmhtl}, and the potential in \Eq{V}. This is also the case if $T\gg \mu$. 
Then the leading $\mu$ dependence is obtained by expanding $m_D$ in $\mu^2/T^2$. Note that in these cases the imaginary part of the potential is parametrically larger than the real part if $m_D \sim 1/r\sim p$, hence bound states can only exist if $1/r \gg m_D$, and cease to exist when this imaginary part takes over the real part, that is before the screening mechanism $r\sim m_D$ is sizable \cite{Escobedo:2008sy}. Even if heavy quarkonium cannot be considered a bound state anymore, its spectral function can be calculated from the evolution by its non-hermitian Hamiltonian \cite{Burnier:2007qm}.

\subsection{Small $T$ ($T\ll \mu$)} 
The case $\mu \gg T$, however, deserves a separate discussion. First of all, the hard contribution from Fig. \ref{1lmm} should be dropped since $T$ is not hard anymore.
When $T\to 0$ the imaginary part of the potential \Eq{V} and mass shift \Eq{dmhtl} vanish, and one may naively think that the bound state is stable. But this need not be so. On the one hand, there could be subleading contributions that do not vanish in this limit, and on the other hand, before $T$ reaches zero, there are additional scales that play a role, in particular the binding energy $E$. Let us analyze the following two cases separately, $p\sim m_D \gg T \gg E$ and 
$p\sim m_D \gg E \gg T$. The case $T\gg p\sim m_D \gg E$ reduces to expanding the results for the $T\sim \mu$ case in $T/\mu$.

\subsubsection{$ T \gg E$ }
\label{mumDTE}

At the scale $m_D$ we still get the same result as in \Eq{V}. The $T$ factor in the imaginary part comes from the Bose enhancement in the gluon distribution function $n_B(k^0)\sim T/k^0$ for $k^0\ll T$, which still holds since $k^0\sim E$, the typical energy transfer, and $E\ll T$. However, now the imaginary part of the potential is parametrically smaller than the real part, and one may wonder whether $T$-independent contributions to the imaginary part exist that compete in size with \Eq{V}. The leading $T$-independent contributions to the imaginary part of the mass shift come from Fig. \ref{2lm} (hard scale) and 
\Fig{fig:mDself1long} ($m_D$ scale), when the internal heavy quark line is 
on-shell.  They are  ${\cal O}(g^2 m_D^2/m)$. Since one-loop contributions to the potential are at most ${\cal O}(g^4 m_D)$, then
these contributions are  
parametrically smaller than the imaginary part of \Eq{dmhtl} and \Eq{V}.

\begin{figure}{
\includegraphics[width=.3\textwidth,angle=0]{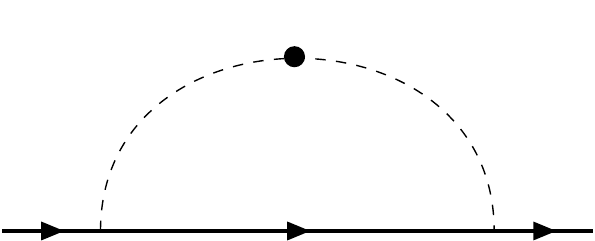}
\caption{Heavy-quark self-energy contribution at the $m_D$ scale. The solid lines are heavy-quark propagators and the dashed line with a dot the HTL longitudinal gluon propagator.
 \label{fig:mDself1long}}}
\end{figure}

The $T$ dependence from the hard scale is encoded in $m_D$. The leading $T$ dependence in the real part of \Eq{V} is $\sim ET^2/\mu^2$. Since $\mu$ is the largest scale in the problem, one may wonder whether other contributions from lower scales are larger. In order to address this question we must take into account that below the $m_D$ scale the only low energy degrees of freedom in the light sector are
the quasi-static magnetic gluons \Eq{magnetic}. 
Furthermore, below the scale $p\sim 1/r$ we can use  
pNRQCD with the mass shifts and singlet potential given in \Eq{dmhtl} and \Eq{V} respectively, and similar modifications to the octet potential, 
\be
V_o= \frac{\left(\frac{C_A}{2}-C_F\right)\als}{r} e^{-m_D r} -C_F\als m_D  \,.
\label{Vom_D}
\ee
At the scale $T$, there is a contribution from Fig. \ref{mg}, in which singlet and octet propagators must be understood  with the potentials described above, 
\be
\delta V=-\frac{\Gamma (\frac{4}{3}) \zeta (\frac{4}{3})}{9\pi^2 \cos{\frac{\pi}{6}}} g^2 C_F r^i (h_o-E) r^i T \left(\frac{\pi m_D^2 T}{4}\right)^\frac{1}{3} \,.
\label{MT}  
\ee
This contribution is ${\cal O}(\als r^2 ET(T m_D^2)^{1/3})$ and hence parametrically larger than $\sim ET^2/\mu^2$ (recall that $m_D \sim p \sim 1/r \sim m\als$ implies that $\mu \sim g m$).
Then the leading $T$-dependence to the energy shift is given by the expectation value of the expression above 
and the decay width by minus twice the expectation value of the imaginary part of  \Eq{V}, 
\be
\delta E_{nl}=-\frac{\Gamma (\frac{4}{3}) \zeta (\frac{4}{3})}{9\pi^2 \cos{\frac{\pi}{6}}} g^2 C_F \left(\frac{N_c\als}{2}\left< re^{-m_Dr}\right>_{nl}+\frac{3}{m}\right) T\left(\frac{\pi m_D^2 T}{4}\right)^\frac{1}{3}\,,
\ee
\be
\Gamma_{nl}=2C_F\als T\left(1-\left<\frac{2}{rm_D}\int_0^\infty dx \,\frac{\sin(m_Dr\,x)}{(x^2+1)^2}\right>_{nl}\right)
\,.
\ee
The expectation values above are calculated with the real part of \Eq{V} in the Hamiltonian.
We have used that $r^i (h_o-E) r^i=\frac{N_c\als}{2} re^{-m_Dr}+\frac{3}{m}$ on physical states in \Eq{MT}.

Let us finally mention that there are parametrically larger $T$-dependent contributions to the mass shift, ${\cal O}(g^2 m_D^2/m)$ from the one-loop self-energy diagram in the region $k\sim m_D$ and $k_0 \sim T$. However these contributions are logarithmic in $T$ and hence very smooth. In addition, they are difficult to calculate. We have displayed in Appendix \ref{loglog} the logarithmically enhanced
contributions. 
There are similar subleading contributions ($\sim g^4 E$) from the region $k_0\sim T$, $k\sim \mu$ that
 in some particular cases may compete with \Eq{MT} as well.

\subsubsection{
 $E \gg T$ }
\label{TggE}

In this case, the imaginary part of the tree-level potential turns out to be zero. This is because all relevant scales are bigger than $T$ and hence $N_B(k^0) \sim$ sgn$(k^0)$. 
Then the imaginary part of the tree-level potential becomes proportional to the absolute value of the transfer energy, which is zero for on-shell heavy quarks in the center of mass frame. 
Regarding the contribution from the heavy quark selfenergy,
one can then work along the same lines as in Sec. \ref{sec:EgtrTlow}  in order to prove that the imaginary part vanishes at one loop, as mass-shift contributions 
 would 
be of the same form as \Eq{eq:deltavLowT2}, now with $a = E - k^2/2m$ (and similarly for the transverse contribution). 
  The leading corrections to the imaginary part may arise from the vertex correction (fig. \ref{fig:vertex1}), the two gluon exchange diagrams (fig. \ref{fig:fourpt}) and the two-loop heavy quark selfenergy (fig \ref{2lqself}). 
We prove in Appendix \ref{2lse} that they also vanish. Therefore the imaginary part of the potential and of the mass shift vanish at leading order and including ${\cal O}(\als)$ corrections.

\begin{figure}[h]{
\includegraphics[width=.15\textwidth]{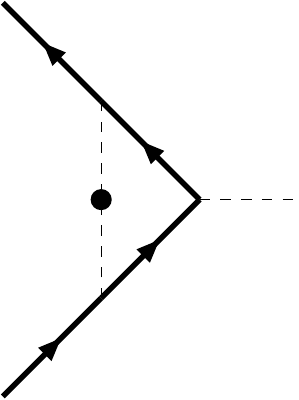}
\caption{Vertex correction at the $m_D$ scale\label{fig:vertex1}. The solid lines are heavy-quark propagators and the dashed line with a dot the HTL longitudinal gluon propagator.}}
\end{figure}

\begin{figure}[h]{
\begin{tabular}{lcr}
\includegraphics[width=.25\textwidth,scale=0.6]{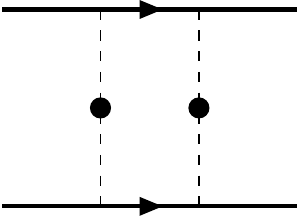}& $\quad\quad\quad$& \includegraphics[width=.25\textwidth,scale=0.6]{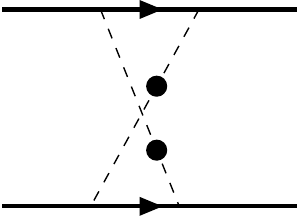}
\end{tabular}
\caption{Two-gluon exchange diagrams. The solid lines are heavy-quark propagators and the dashed lines with a dot HTL longitudinal gluon propagators. \label{fig:fourpt}}}
\label{2ge}
\end{figure}

\begin{figure}[h]{
\begin{tabular}{lcr}
\includegraphics[width=.25\textwidth]{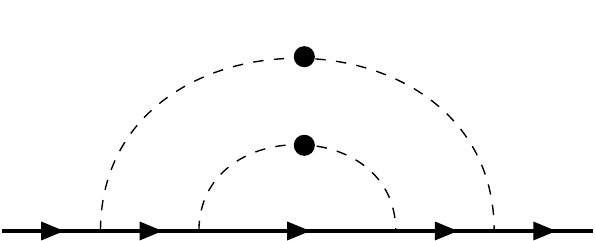}&$\quad\quad\quad$&\includegraphics[width=.25\textwidth]{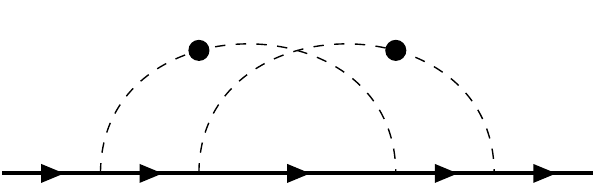}
\end{tabular}
\caption{Two-loop contributions to the heavy-quark selfenergy. The solid lines are heavy-quark propagators and the dashed lines with a dot HTL longitudinal gluon propagators.  \label{2lqself}}}
\end{figure}

Let us next focus on the temperature dependence of the energy shift. The potential depends on temperature through the Debye mass, which gives a $T$-dependent contribution to the energy shift of ${\cal O}(g^2 m_D T^2/\mu^2)$. Since $\mu$ is the largest scale in the problem after the heavy quark mass, we may expect more important contributions from lower scales. We find that the leading $T$-dependent contribution comes from the one-loop self-energy diagram in which the longitudinal gluon propagator has $k\sim m_D$ and $k_0\sim T$, which is ${\cal O}(g^2T^2/E)$. This contribution is difficult to calculate. On the one hand the energy scale $k_0\sim T \ll E$, and hence bound state effects cannot be ignored. On the other hand, pNRQCD cannot be straightforwardly used since $k\sim m_D \sim p \sim 1/r$, and hence the multipole expansion does not hold. In order to avoid the last problem we shall restrict ourselves to the particular case $m\gg \mu \gg p\gg m_D \gg E \gg T$. The $T$-dependent part of the energy shift can be obtained from the second term in \Eq{deltaVR}, where now $a=E-h_o-k^2/4m$. Notice that we have included the quarkonium center of mass recoil energy, which was negligible in Sec. \ref{pTmuE}. For $k\sim m_D$ and $k_0\sim T$ the HTL propagator must be used. We obtain,
\be
\label{mmupmDETleadingTmD}
\delta E_{nl}\vert_{k\sim m_D}=\frac{g^2C_FT^2m_D^2}{36}\left< r^i\frac{1}{E-h_o}r^i\right>_{nl}\left(-\frac{1}{\epsilon}-\frac{1}{3}-2\log \frac{2m_D}{\nu} + \log \pi +\psi\left(\frac{3}{2}\right) \right)\,.
\ee
The $1/\epsilon$ arises from an UV divergence in $k$. It should be compensated by an IR divergence of a contribution at a higher
$k$ scale. In \Eq{deltaVR}, the scale $k\sim \sqrt{-4m(E-h_o)}\sim p\gg m_D$ is also relevant. It allows to make an expansion in $m_D$ in the HTL propagators, which induces the IR divergence we are looking for. We obtain,
\be
\label{mmupmDETleadingTp}
\delta E_{nl}\vert_{k\sim p}=\frac{g^2C_FT^2m_D^2}{36}\left< r^i\frac{1}{E-h_o}\left(\frac{1}{\epsilon}-\frac{2}{3}+2\log \frac{h_o-E}{\nu} - \log 4\pi -\psi\left(\frac{3}{2}\right) \right)r^i\right>_{nl}\,.
\ee
There is a problem with the result above: for $k\sim p$ the multipole expansion, on which pNRQCD is based, does not hold. Nevertheless, if we are only interested in the IR behaviour $k\to 0$, namely $k\ll p$, then it can be used. That means that our calculation above gets the correct IR behaviour, and hence the correct log, but the finite pieces are not reliable. Putting together \Eq{mmupmDETleadingTmD} and \Eq{mmupmDETleadingTp}, we then obtain,
\be
\delta E_{nl} =-\frac{g^2C_FT^2m_D^2}{18}\left< r^i\frac{1}{E-h_o}\left(\log \frac{m_D}{h_o-E} +{\cal O}(1)  \right)r^i\right>_{nl}\,,
\ee
where the ${\cal O}(1)$ means there is an unknown number that adds to the logarithmically enhanced contribution.

\section{Discussion}
\label{disc}

We have worked out the modifications in the binding energy and decay width that a QGP at high temperature and/or chemical potential
induces in a heavy quarkonium state, generalizing earlier work done in the limit of a vanishing chemical potential. 
This was done from QCD at weak coupling in the real-time formalism with approximations that are well under control, relying on the hierarchy of scales in the problem. This is in contrast with earlier work on heavy quarkonium at finite chemical potential, in which some modeling is introduced \cite{Kakade:2015laa}. In particular, we have shown that the rather usual assumption that the medium effects can be encoded in a modified potential, as made in the early days \cite{Gao:1996xz,Liu:1997tc}, is not always true. {Note that this is independent on whether the models fit well lattice results on the potential, like, for instance, refs. \cite{Burnier:2015nsa,Guo:2018vwy}, since lattice potentials do not encode non-potential effects either. Non-potential effects require the full quarkonium dynamics and not just static quarks \cite{Voloshin:1979uv,Leutwyler:1980tn}.}

We have restricted ourselves to heavy quarkonia at rest. The effects of a relative velocity with respect to the thermal bath
may eventually be addressed along the lines of refs. \cite{Escobedo:2011ie,Escobedo:2013tca}. In fact, when the effects of the medium can entirely be encoded in a potential, they have already been addressed in \cite{Thakur:2016cki}. We have focused on a number of cases in which analytic results can be produced. However, it should be clear from our general formulas that numerical results can be also obtained for the remaining cases.

When the temperature and chemical potential are smaller than the typical momentum exchange between the heavy quarks, the medium effects are a perturbation that, in general, cannot be encoded in a potential. This has been already emphasized for zero chemical potential in \cite{Escobedo:2008sy,Brambilla:2008cx,Brambilla:2010vq,Escobedo:2010tu}. In this case, Coulomb resummations must be always carried out, and the medium effects enter through
gluons emitted by chromoelectric dipole transitions which turn a color-singlet quarkonium into a color-octet one or viceversa. In that respect it is very helpful to use pNRQCD. Depending on the energy and momentum of the emitted gluon, HTL resummations may also be necessary. If the chemical potential and the temperature have the same size, the results we obtain are similar to the ones of the zero chemical potential case, but include non-trivial functions of $\mu/T$. If $T\gg\mu$, we can just expand our results in $\mu/T$, as the distribution functions are analytic in $\mu/T$. However, if $\mu\gg T$, the distribution functions are not analytic in $T/\mu$, and this requires extra care.
In this limit, the 
 Debye mass $m_D\sim g\mu$ may be comparable to $T$ and hence accounting properly for the leading temperature effects requires HTL resummations.  We find that if the temperature is larger than the binding energy, the decay width is proportional to $T$, but it vanishes otherwise.

When the temperature or the chemical potential are larger than the typical momentum exchange between the heavy quarks, the medium effects modify the leading order potential. This is the case addressed in the pioneering works \cite{Matsui:1986dk}, in which the screening was proposed as the mechanism leading to $J/\psi$ suppression. Later on, an important imaginary part due to Landau damping was uncovered for this potential which changed the picture \cite{Laine:2006ns}. When $T\simeq \mu$ the imaginary part of the potential is proportional to $g^2T$ and parametrically larger than the real part ($\sim g^2 m_D$). Due to this imaginary part, the heavy quarkonium melts before noticing the screening effects, as in the case of zero chemical potential \cite{Escobedo:2008sy}. When $T\simeq m_D\sim g\mu$, screening and Landau damping compete for being the leading effect. The imaginary part of the potential exists as long as the temperature is larger than the binding energy, but it vanishes otherwise. We have been able to prove it at next-to-leading order in $\als$.

Our analysis turned out to be technically challenging,
 as Coulomb and/or HTL resummations have been necessary in several instances. The use of effective field theories has been invaluable to keep track of the important terms in a systematic manner. Dimensional regularization has been used to regulate both the IR and UV divergencies that arise in the intermediate steps of the calculations when we factorize the contributions of the different scales. 
 We have obtained contributions from energy and momentum regions that had been ignored so far. In that respect the method of integration by regions developed in \cite{Beneke:1997zp} (see \cite{Smirnov:2002pj} for a review) has also been very useful. For instance, in order to get the leading temperature effects in the binding energy when $\mu \gg p \sim m_D\gg E\gg T$ we needed gluons of energy $\sim T$ and momentum $\sim m_D$. These gluons are on the one hand sensitive to the binding energy, and hence Coulomb resummations are required, and on the other hand have a momentum large enough so that the multipole expansion cannot be applied, and hence the calculation cannot be carried out entirely in pNRQCD. We circumvented these difficulties by introducing the extra hypothesis $p\gg m_D$. Another non-trivial example is the contribution of quasistatic magnetic modes \cite{Linde:1980ts,Gross:1980br} when $\mu \gg p \sim m_D\gg T\gg E$ that give an important $T$-dependent piece of the binding energy\footnote{Quasistatic magnetic modes are the responsible for perturbation theory at finite temperature to break down at energy scales smaller than the Debye mass $k_0\ll m_D\sim gT$. This is due to Bose enhancement that introduces large factors $T/k_0$ in the thermal propagators, which compensate for the $g$s in the vertices. Note that here the situation is different. Since $m_D\sim g\mu\gg T\sim k_0$, there is no Bose enhancement and perturbation theory is well under control.}. Finally, let us mention the logarithmic $T$-dependence in the same case, which is log enhanced and requires the introduction of an extra regularization to factorize the energy scale from the momentum scale. We have chosen an analytic regularization similar to ref. \cite{Becher:2011dz}, see Appendix \ref{loglog}.

Our results are obtained entirely in the weak coupling regime of QCD and thus may not be straightforwardly applied to realistic experimental situations, especially for charmonium, as some of the scales in the problem may not be large enough. Nevertheless,
we believe they provide important constraints to models, as they fix quite a number of asymptotic behaviours for large $\mu$ of more realistic models. In the absence of a definitive approach to address real-time phenomena in general and large chemical potentials in particular in lattice QCD, complementary approaches based on weak coupling QCD should be helpful. 

Let us then consider $J/\psi$, which will be observed in most of the planned experiments \cite{Galatyuk:2019lcf}. If we take $m_c\sim 1.6$ GeV\footnote{This value corresponds to the so called RS' mass at low scale in ref. \cite{Peset:2018ria}.}, then the experimental value of the $J/\psi$ mass delivers $E\sim 0.1$ GeV. If we associate this value with a Coulombic state, we obtain
$\als (p) \sim 0.4$ and $p\sim 1/r \sim 0.4$ GeV. We see that the value of $p$ is very low even if $\als$ is relatively small \footnote{This in fact means that assuming a Coulombic bound state at leading order is not really consistent. One needs to include higher orders in $\als$ in the potential to get $J/\psi$ under reasonable control, see for instance \cite{Peset:2018ria} and references therein.}.
For the maximum expected values of the baryon chemical potential $\mu_B$ quoted in ref. \cite{Galatyuk:2019lcf}, we have $\mu=\mu_B/3\lesssim 0.3$ GeV\footnote{We understand that the units for $\mu_B$ in table 1 of \cite{Galatyuk:2019lcf} are MeV rather than the quoted GeV.}. It means that most of the times we would be in the case of Sec. \ref{pTmuE}, and only when $\mu \sim 0.3$ GeV, the case of Sec. \ref{TmupE} may be relevant. This is of course provided that $T\lesssim \mu$. 

Although analyzing bottomonium does not seem to be in the future experimental plans, some of the colliders feeding the relevant experiments (e.g. NICA, RHIC, SPS) are energetic enough to produce it. If we take $m_b\sim 4.9$ GeV\footnote{This value corresponds to the so called RS' mass at low scale in ref. \cite{Peset:2018ria}.}, then the experimental value of the $\Upsilon (1S)$ mass delivers $E\sim 0.34$ GeV. If we associate this value with a Coulombic state, we obtain
$\als (p) \sim 0.4$ and $p\sim 1/r \sim 1.2$ GeV. Then, for the expected values of the chemical potential, $\Upsilon (1S)$ would always be in the case of Sec. \ref{pTmuE}. 

If we stick to qualitative features of our results, the most relevant one is that a temperature larger than the size of the binding energy $T>E$ appears to be necessary for heavy quarkonium to develop a decay width. No decay width is developed if $T<E$ , no matter how large is the chemical potential (provided it is smaller than the heavy quark mass). This may be understood in terms of the Fermi sea: In order to dissociate quarkonium, a light quark of the Fermi sea must provide an energy larger than the binding energy to the bound state. But then it becomes less energetic in the final state, and since all the states with less energy are occupied in the Fermi sea, the process cannot take place. Hence at large chemical potential and small temperature, we only expect modifications in the heavy quarkonium mass (through the binding energy). The dissociation mechanism would be screening, namely the one originally proposed in \cite{Matsui:1986dk}.

 For sufficiently heavy quark mass, chemical potential and/or temperature, our results are reliable. In the case of small (zero) temperature and large chemical potential, one should observe in the quarkonium spectral function a shift in the location of each bound state peak with no modifications in the width when we increase $\mu$. This is in contrast with what happens at large temperature and small (zero) chemical potential, in which case, apart from the shift in the location of the bound state peaks, a widening 
of the peaks is observed when the temperature is increased. In fact, the melting of the bound states occurs because the peaks corresponding to different bound states overlap and lose their identity. This can be understood at weak coupling in terms of the Landau damping \cite{Laine:2006ns}. In the case of large chemical potential one would just observe bound states peaks disappearing when we increase the chemical potential.
It would be interesting to cross-check our results in lattice QCD simulations, but this would require having overcome the difficulties of dealing with a large chemical potential (see \cite{Aarts:2013bla,Aarts:2015tyj,Gattringer:2016kco,Aarts:2016hap,Ratti:2018ksb,Banuls:2019rao} for reviews).

However, we can compare with the results of ref. \cite{Hands:2012yy}, a NRQCD lattice simulation for $N_c=2$ and $N_f=2$ and heavy quark mass $m a=5,4,3$, where $a$ is the lattice spacing. They consider $0\leq \mu a\leq 1.1$ and $1/24\leq T a \leq 1/12$, hence we can probe the $\mu \gg T$ regime. If we assume that the binding energies are Coulombic, from the values for different masses of $\Delta E a$ at $\mu=0$ in their Fig. 1, we obtain that $\als\sim 0.65-0.7$ at the scale of the typical relative momentum $p$. This implies $E\sim 0.3 a$ and $p\sim 1/r\sim 1.2 a$. Hence, most of the data displayed in their Fig. 1 is in the region $m\gg p\gg \mu \gg E \gg T$, and we should compare it with the results in our \Eq{deltaEmDTE} and \Eq{pmuETTdep}. The left panel of their Fig. 1 shows the binding energy as a function of $\mu a$ for three values of the heavy quark mass. For these plots to be compatible with \Eq{deltaEmDTE}, we need the total (i.e. including the one hidden in the Debye mass) coefficient of the $\mu^3$ term to be positive (the temperature can be neglected). This is achieved if $\als (m_D)\gtrsim 0.86$. If so, our expression qualitatively describes the rising observed from $\mu a\sim 0.6$ to $\mu a\sim 1$. We can also understand the bending downwards around $\mu a\sim 1$: in this region $\mu \sim p$ and with our values of $\als (m_D)$,
$\mu\sim m_D$, hence \Eq{dmhtl} and \Eq{V} should better be used for the energy shift. If we expand \Eq{V} for $m_D r \ll 1$,
the first correction to the Coulomb potential is negative, which may explain the above mentioned downward trend. However, we cannot explain 
the mild decreasing from $\mu a \sim 0.3$ to $\mu a \sim 0.6$. We would probably need expressions for $\mu \sim E$ that we have not worked out, or it may simply happen that $\als$ becomes too large at those low scales so that our weak coupling description is not appropriate even qualitatively. In any case, the behavior of the curves with the mass is easy to understand as the dependence on the chemical potential goes as $\mu^3/m^2$. Hence the smaller the mass is, the more noticeable the effects are, as clearly shown in the left panel. The temperature effects are displayed in the right panel of their Fig. 1. Those should be encoded in \Eq{pmuETTdep}, and we indeed see in the plot that rising the temperature increases the binding energy, although we do not observe the quadratic increase of \Eq{pmuETTdep}.
     
Finally, our results can also be applied to the case of a non-vanishing isospin chemical potential $\mu_I$ rather than a baryon chemical potential. We only have to replace $N_f \mu$ by $2\vert \mu_I\vert$ in our equations. This is because our expressions are symmetric under $\mu \leftrightarrow -\mu$ and each light quark contributes the same amount of $\mu$ at finite baryon chemical potential. At finite isospin chemical potential, the $u$-quark contributes by $\mu_I$, the $d$-quark by $-\mu_I$. We could then try to compare with the two-flavor lattice results of ref. \cite{Detmold:2012pi}. However, the results displayed in that reference correspond to $\mu_I\lesssim 0.3$ GeV, a too low scale to apply our weak coupling calculation\footnote{In addition, they sit in the region $\mu_I \lesssim E$ for which we do not have explicit formulas.}. 

\acknowledgments

We have been supported by the MINECO (Spain) under the projects FPA2016-76005-C2-1-P and PID2019-105614GB-C21, and by the 2017-SGR-929  grant (Catalonia). J.S. has also been supported by the projects FPA2016-81114-P and PID2019-110165GB-I00 (Spain). 
We also acknowledge financial support from the State Agency for Research of the Spanish Ministry of Science and Innovation through the “Unit of Excellence Mar\'ia de Maeztu 2020-2023” award to the Institute of Cosmos Sciences (CEX2019-000918-M).

\appendix

\section{Leading corrections to the imaginary part of the potential and heavy quark selfenergy in the $\mu \gg p\sim m_D \gg E \gg T$ case}\label{2lse}

We prove in this Appendix that the leading corrections to the imaginary part of the potential and the heavy quark selfenergy in Sec. \ref{TggE} also vanish.

\subsection{HTL correction to the vertex}

Beyond leading order, a possible source of imaginary contributions to the potential is the vertex correction of \Fig{fig:vertex1}. Writing the full vertex function as $W^a=T^a (1+\delta W)$,  $\delta W$ reads
\begin{align}
i\delta W & = -\frac{(ig^3)}{2N_c} \nu^{4-D} \int \frac{d^Dk}{(2\pi)^D}
\frac{1}{\tilde E -k^0 +i\eta} \frac{1}{\tilde E' -k^0 +i\eta} 
D^{}_{00}(k_0,k)\,,
\end{align}
where $\tilde E=E-(\mathbf{k}+\mathbf{p})^2/(2m)$ and $\tilde E'=E'-(\mathbf{k}+\mathbf{p}')^2/(2m)$, $(E,\mathbf{p})$ and $(E',\mathbf{p}')$ being the incoming and outgoing heavy-quark energy and three-momentum respectively. Eventually, we will use that
$\tilde E$, $\tilde E'$ are much smaller than $k$, $p$ and $p'$, and that in a bound state $E$, $E' <0$. However this limit must be taken once the $k_0$ integral has been carried out, otherwise we are left with ill-defined expressions. Since in the small-temperature limit (here and in the following we will use the shorthand notation $D_{00}^{R/A}(k^0,\k) \equiv R/A(k^0,k)$ and $\dot{R}(k^0,\k) = dR(k^0,k)/dk^0$)
\be
D^{}_{00}(k_0,k) =\frac{1}{2}\Big[R(k_0,k)+A(k_0,k)+{\rm sgn}(k_0)\left(R(k_0,k)-A(k_0,k)\right)\Big]\,,
\ee
we write $\delta W=\delta W_1+\delta W_2$, with $\delta W_2$ containing the terms proportional to sgn$(k_0)$ and $\delta W_1$ all the rest.
$\delta W_1$ can be evaluated by contour integration, and the small $\tilde E$, $\tilde E'$ limit gives,
\be
\delta W_1=-\frac{(ig^3)}{4N_c} \nu^{4-D} \int \frac{d^dk}{(2\pi)^d}\dot{R}(0,k)\,,
\label{w1}
\ee
which is purely imaginary. The imaginary part of $\delta W_2$ can also be evaluated using the formula,
\be
\frac{1}{\tilde E -k^0 +i\eta} \frac{1}{\tilde E' -k^0 +i\eta} =\frac{{\cal P}}{\tilde E -k^0 } \frac{{\cal P}}{\tilde E' -k^0 }
 -i\pi \delta (\tilde E -k^0 ) \frac{{\cal P}}{\tilde E' -k^0 } -i\pi
\frac{{\cal P}}{\tilde E -k^0 } \delta(\tilde E' -k^0 )\,. 
\ee
It leads to
\be 
{\rm Im}\, \delta W_2=\frac{(ig^3)}{4N_c} \nu^{4-D} \int \frac{d^dk}{(2\pi)^d}\dot{R}(0,k)\,,
\ee
which cancels exactly (\ref{w1}). Hence no imaginary part arises from the vertex correction at one loop.

\subsection{HTL two-gluon exchange contributions to the potential}

The two gluon exchange contributions of Fig. \ref{fig:fourpt} may also provide imaginary parts. Consider first the diagram on the left projected on color singlet states. This diagram contains the iteration of the leading order potential, which must be subtracted,
\be
\delta V=-iC_F^2\int\frac{d^Dk}{(2\pi)^D}\frac{i}{\frac{E}{2}+k_0+i\eta}\frac{i}{\frac{E}{2}+k_0+i\eta}\Big[ D_{00}(k_0,\mathbf{k})D_{00}(k_0,\mathbf{k}+\mathbf{q})-D_{00}(0,\mathbf{k})D_{00}(0,\mathbf{k}+\mathbf{q})\Big]\,,
\ee
where we recall that $D_{00}$ stands for the $11$ component of the real-time temporal gluon propagator. Upon writing it in terms of the 
retarded and advanced propagators 
 we obtain
\bea
\delta V &=&\delta V_1 +\delta V_2  \,, \quad {\rm with} \nn\\
\delta V_1 &=&-iC_F^2\int\frac{d^D k}{(2\pi)^D}\frac{i}{\frac{E}{2}+k_0+i\eta}\frac{i}{\frac{E}{2}+k_0+i\eta}\frac{1}{2}\Big[R(k_0,\mathbf{k})R(k_0,\mathbf{k}+\mathbf{q})-R(0,\mathbf{k})R(0,\mathbf{k}+\mathbf{q})\\\nn\\
&& + A(k_0,\mathbf{k})A(k_0,\mathbf{k}+\mathbf{q})-A(0,\mathbf{k})A(0,\mathbf{k}+\mathbf{q})\Big] \,, \\
\delta V_2 &=&-iC_F^2\int\frac{d^D k}{(2\pi)^D}\frac{i}{\frac{E}{2}+k_0+i\eta}\frac{i}{\frac{E}{2}+k_0+i\eta}\frac{1}{2}{\rm sgn}(k^0)\Big[R(k_0,\mathbf{k})R(k_0,\mathbf{k}+\mathbf{q})-A(k_0,\mathbf{k})A(k_0,\mathbf{k}+\mathbf{q})\Big]\,.
\eea
For $\delta V_1$  the integral over $k_0$ can be carried out,
which turns the two heavy quark propagators into a $i/(E+i\eta)$ quarkonium propagator and replaces the $k_0$ in the retarded and advanced propagators by $E/2$ and $-E/2$ respectively. Using $E\ll  \mathbf{k}\,,\mathbf{q}$, we finally get 
\be
{\rm Im}(\delta V_1)=-\frac{iC_F^2}{2}\int\frac{d^d k}{(2\pi)^d}\left(R(0,\mathbf{k})\dot{R}(0,\mathbf{k}+\mathbf{q})+\dot{R}(0,\mathbf{k})R(0,\mathbf{k}+\mathbf{q})
\right)\,,
\label{2gedirect}
\ee
where we have also used $R(0,\mathbf{k})=A(0,\mathbf{k})$ and $\dot{R}(0,\mathbf{k})=-\dot{A}(0,\mathbf{k})$. For $\delta V_2$, it is tempting to take $E\to 0$ in the integrand, and then formally show that Im$(\delta V_2)=0$. However, it turns out that the integral is ill-defined in that limit, and this naive result is wrong. Instead, one can show that, for $E<0$,
\bea\label{2gedirect2}
{\rm Im}(\delta V_2)&=&-\frac{iC_F^2}{4E}\int\frac{d^d k}{(2\pi)^d}\left(R(-\frac{E}{2},\mathbf{k})R(-\frac{E}{2},\mathbf{k}+\mathbf{q})-A(-\frac{E}{2},\mathbf{k})A(-\frac{E}{2},\mathbf{k}+\mathbf{q})\right.\nn\\
&&\left. -R(\frac{E}{2},\mathbf{k})R(\frac{E}{2},\mathbf{k}+\mathbf{q})+A(\frac{E}{2},\mathbf{k})A(\frac{E}{2},\mathbf{k}+\mathbf{q})\right)\\
&=& \frac{iC_F^2}{2}\int\frac{d^d k}{(2\pi)^d}\left(R(0,\mathbf{k})\dot{R}(0,\mathbf{k}+\mathbf{q})+\dot{R}(0,\mathbf{k})R(0,\mathbf{k}+\mathbf{q})\right)\,,\nn
\eea
where we have used $E\ll  \mathbf{k}\,,\mathbf{q}$ in the last equality. Note that (\ref{2gedirect2}) cancels exactly (\ref{2gedirect}), so that finally ${\rm Im}(\delta V)={\rm Im}(\delta V_1)+ {\rm Im}(\delta V_2)=0$.

For the diagram on the right of Fig. \ref{fig:fourpt} we get, in a similar way, cancellations between the imaginary part of the terms proportional to sgn$(k_0)$ and the rest of the contribution. Then the one-loop contribution to the imaginary part of the potential also cancels out.

\subsection{Two-loop HTL contributions to the heavy quark self-energy}

At the same order, namely suppressed by $\als$, there are also the two loop contributions to the heavy quark self-energy. 
We have three diagrams contributing to the heavy quark self-energy at two loops. 
One of the diagrams corresponds to a longitudinal HTL gluon self-energy insertion to the one loop diagram. The imaginary part of this diagram has been shown to vanish on general grounds in Sec. \ref{TggE} . We show in the following sections that the remaining two diagrams, which are shown in Fig. \ref{2lqself},  also have a vanishing imaginary part.

\subsubsection{Heavy quark self-energy insertion}

The diagram on the left of Fig. \ref{2lqself} corresponds to a heavy quark selfenergy insertion to the one-loop diagram. The (complex) mass shift produced by this diagram reads
\bea
\delta m &=& ig^2C_F \nu^{4-D} \int \frac{d^Dk}{(2\pi)^D} \frac{i}{(E - k_0 -\Sigma (E-k_0) +i\eta)}     D^{HTL}_{00}(k_0,k) \,, \quad {\rm with} \label{hqse2l}\\
\Sigma (E-k_0,\boldmath{k}) &=& ig^2C_F \nu^{4-D} \int \frac{d^Dk'}{(2\pi)^D} \frac{i}{(E - k_0 -k_0' +i\eta)}     D^{HTL}_{00}(k_0',k')  \,.
\eea

The imaginary part of the one-loop heavy quark selfenergy reads,
\be
{\rm Im} \Sigma (E-k_0)= \frac{g^2C_F}{2}\nu^{4-D}\theta (E-k_0)\int \frac{d^dk}{(2\pi)^d}\left(R(E-k_0,k)-A(E-k_0,k)\right)\,.
\ee
Note that in (\ref{hqse2l}) $\Sigma$ is a perturbation, and hence it can only slightly move the location of the pole.
Near this location we can then use that  Im $[\Sigma (0)]=0$ (since $R(0,k)=A(0,k))$, thus the heavy quark propagator pole will still be in the upper complex half-plane. This observation allows to calculate,
\bea
{\rm Im} (\delta m) &=&\frac{ig^2C_F}{2}\nu^{4-D} \int\frac{d^Dk}{(2\pi)^D}\Big[{\cal P}\frac{1}{(E - k_0 -\Sigma (E-k_0) )} \left(R(k_0,k)+A(k_0,k)\right)\nn\\
&& -\pi i \delta \left(E - k_0 -\Sigma (E-k_0) \right) {\rm sgn}(k_0) \left(R(k_0,k)-A(k_0,k)\right)\Big]\nn\\
&=&\frac{g^2C_F\pi}{4}\nu^{4-D} \int \frac{d^dk}{(2\pi)^d}\Big[ \left(R(E-\Sigma (0),k)+A(E-\Sigma (0),k)\right) \\
&& +  {\rm sgn}(E-\Sigma (0)) \left(R(E-\Sigma (0),k)-A(E-\Sigma (0),k)\right)\Big]=0\,,\nn
\eea
where in the last equality we have used that ${\rm sgn}(E-\Sigma (0))=-1$, since $\Sigma (0)$ is a perturbation and $E<0$ for a bound state.

\subsubsection{The irreducible diagram}

We focus here on the diagram on the right of Fig. \ref{2lqself}. We have,
\bea
\delta m &=&\frac{iC_Fg^4}{16 N_c}\nu^{8-2D}\int  \frac{d^Dk}{(2\pi)^D} \frac{d^Dk'}{(2\pi)^D} \frac{i}{(E  -k_0' +i\eta)}\frac{i}{(E - k_0 -k_0' +i\eta)}\frac{i}{(E - k_0  +i\eta)}\nn\\
&& \Bigg\{ \Big(R(k_0,k)+A(k_0,k)\Big)\Big(R(k_0',k')+A(k_0',k')\Big) +{\rm sgn}(k_0) \Big(R(k_0,k)-A(k_0,k)\Big){\rm sgn}(k_0') \Big(R(k_0',k')-A(k_0',k')\Big)\nn\\
&& +\Big(R(k_0,k)+A(k_0,k)\Big) {\rm sgn}(k_0') \Big(R(k_0',k')-A(k_0',k')\Big)+{\rm sgn}(k_0) \Big(R(k_0,k)-A(k_0,k)\Big)\Big(R(k_0',k')+A(k_0',k')\Big) \Bigg\}\nn \\
&\equiv& \delta m_1+\delta m_2+2\delta m_3 \,, 
\eea
where in the definitions above we have used that the two terms in the third row are equivalent. For $\delta m_1$, the  $k_0'$ integral can be done by contour integration, and the limit $E\ll k, k'$ is well defined. We obtain
\bea
\delta m_1&=&-\frac{iC_F g^4}{16 N_c}\nu^{8-2D} \int\frac{d^dk}{(2\pi)^d} \frac{d^dk'}{(2\pi)^d}\left\lbrace \int_{-\infty}^\infty \frac{dk_0}{(2\pi)}\frac{R(k_0,k) A(k_0,k')}{(k_0  - i\eta)^2}
-i\dot{R}(0,k)R(0,k')\right\rbrace\,,
\eea
where we have dropped terms with two advanced propagators as all singularities are in the upper half plane. For $\delta m_2$, the limit $E\ll k, k'$ is also well defined due to the fact that $R(0,k)=A(0,k)$. We obtain,
\bea
\delta m_2&=&-\frac{iC_F g^4}{16 N_c}\nu^{8-2D}\int \frac{d^dk}{(2\pi)^d} \frac{d^dk'}{(2\pi)^d}\int_0^\infty\frac{dk_0}{(2\pi)}\int_0^\infty dk_0' \frac{\left(R(k_0,k)-A(k_0,k)\right) \left(R(k_0',k')-A(k_0',k')\right)}{k_0 k_0'}\left(\delta(k_0+k_0')+\delta(k_0-k_0') \right) \nn\\
&=&-\frac{iC_F g^4}{32 N_c}\nu^{8-2D}\int \frac{d^dk}{(2\pi)^d} \frac{d^dk'}{(2\pi)^d}\int_{-\infty}^\infty\frac{dk_0}{(2\pi) k_0^2}\Big[
\left(R(k_0,k)-A(k_0,k)\right) \left(R(k_0,k')-A(k_0,k')\right)
\Big]\,.
\eea
The $k_0^2$ in the denominator can be substituted by $(k_0-i\eta)^2$. Then the term with two advanced propagators can be dropped, and the two terms with one advanced and one retarded propagator are equivalent (upon $\mathbf{k}\leftrightarrow \mathbf{k}'$)
When we add up $\delta m_1$ and $\delta m_2$, we have,
\bea
&& 
\delta m_1+\delta m_2=-\frac{iC_F g^4}{32 N_c}\nu^{8-2D} \int \frac{d^dk}{(2\pi)^d} \frac{d^dk'}{(2\pi)^d}\left[\int_{-\infty}^\infty\frac{dk_0}{(2\pi) }\frac{R(k_0,k)R(k_0,k')}{(k_0-i\eta)^2}
 -i 2 \dot{R}(0,k)R(0,k')\right]=0\,,
\eea
where in the last equality we have evaluated the $k_0$ integral by contour integration and used the symmetry $k\leftrightarrow k'$, which exactly cancels the last term.

Consider finally $\delta m_3$. The integral over $k_0'$ (or $k_0$) can be done by contour integration, then we are left with an expression with a well-defined $E\ll k, k'$ limit,
\be
\delta m_3=\frac{i C_F}{8 N_c}\nu^{8-2D} \int \frac{d^dk}{(2\pi)^d} \frac{d^dk'}{(2\pi)^d}\int_0^\infty\frac{dk_0}{(2\pi)}
\frac{R(k_0,k')-A(k_0,k')}{{k_0}^2}\Big[ R(0,k)+A(0,k)-R(k_0,k)-A(k_0,k)\Big] \,.
\ee
From this expression, it is easy to see that Im $(\delta m_3)=0$ (recall that $R^\ast(k_0,k)=-A(k_0,k)$). Hence, Im$(\delta m)=$Im$(\delta m_1+\delta m_2+2\delta m_3)=0$ at two loop level as well.

\section{$\log T$-dependent log-enhanced mass shift contributions}
\label{loglog}

We mentioned at the end of Sec. \ref{mumDTE}, this is in the case $m\gg \mu\gg m_D\sim p \gg T \gg E$, that there are parametrically larger $T$-dependent contributions than those stemming from the magnetic gluons. They correspond to the region $k^0 \sim T$, $k\sim m_D$ in the self-energy diagram of \Fig{fig:mDself1long}. The $T$-dependent contribution to the (real part of the) mass shift can be obtained from 
\be\label{deltam}
\delta m  =  g^2 C_F 
\nu^{4-D}\int \frac{d^dk}{(2\pi)^d} {\cal P} \int_0^\infty \frac{dk^0}{2\pi} \Big( \frac{1}{a-k^0}+\frac{1}{a+k^0} \Big) \frac{N_B(k^0)}{2}  \Big[ D_{00}^R(k^0,k) - D_{00}^A(k^0,k) \Big] \,,
\ee
where $a=E-k^2/2m$. For $k\sim m_D\sim p$ and $k_0\sim T$, then $k^2/2m \sim E \ll T\sim k_0$, $a$ can be expanded in the denominators, and $k_0$ in the HTL longitudinal gluon propagators. We have,
\be
\delta m\vert_{k\sim m_D}^{k_0\sim T}  \simeq  -g^2 C_F 
\nu^{4-D}\int \frac{d^dk}{(2\pi)^d}  \int_0^\infty \frac{dk^0}{2\pi} \frac{a}{k^0}  N_B(k^0)  \frac{\pi m_D^2}{k(k^2+m_D^2)^2} \ \,.
\ee
The expression above not only contains a UV log divergence in $k$, which is already regulated in DR, but also an IR power divergence and a UV log divergence in $k_0$, which need regularization. We choose the analytic regularization $dk_0\to dk_0 (k_0/\nu')^\lambda$, $\lambda\to 0$. 
This regularization drops the power-like divergences as DR does, and hence we are left with the UV log divergence that will be represented by a pole in $1/\lambda$. We obtain,
 \be\label{mD1olambdaUV}
\delta m\vert_{k\sim m_D}^{k_0\sim T}  =  -g^2 C_F 
\nu^{4-D}\int \frac{d^dk}{(2\pi)^D}    \frac{a\pi m_D^2}{k(k^2+m_D^2)^2} \left(-1/\lambda +\gamma-\log \frac{2\pi T}{\nu'}\right) \,.
\ee
The $1/\lambda$ pole above must be compensated by the IR behavior of $k^0$ at a higher scale, while keeping $k$ at the same size.
A natural choice is taking $k_0\sim k \sim m_D$. Then the same approximations as before can be done in the heavy quark propagator, but the distribution function $N(k_0)$ reduces to $1$ and the longitudinal gluon HTL propagators must be kept exact. We have,
\be
\delta m\vert_{k\sim m_D}^{k_0\sim k}  \simeq  -g^2 C_F 
\nu^{4-D}\int \frac{d^dk}{(2\pi)^d}  \int_0^\infty \frac{dk^0}{2\pi} \Big(\frac{a}{k^0} \Big) \Big[ \frac{\pi m_D^2 }{k} \theta (k-k_0) \vert  D_{00}^R(k^0,k)\vert^2\Big] \,.
\ee
This expression is independent of $T$. We only need it to make sure that the $1/\lambda$ of \Eq{mD1olambdaUV} cancels against the IR behavior of a higher energy contribution. Then we can safely take the $k_0\to 0$ limit in $D_{00}^R(k^0,k)$ above. Upon implementing the
analytical regularization discussed above we obtain
 \be\label{mD1olambdaIR}
\delta m\vert_{k\sim m_D}^{k_0\sim k}  \simeq  -g^2 C_F 
\nu^{4-D}\int \frac{d^dk}{(2\pi)^D}    \frac{a\pi m_D^2}{k(k^2+m_D^2)^2} \left(1/\lambda +\log \frac{k}{\nu'}\right) \,.
\ee
Putting \Eq{mD1olambdaUV} and \Eq{mD1olambdaIR} together, we have the following $T$-dependent contribution,
\be
\label{mDUV}
\delta m\vert_{k\sim m_D}  \simeq  -g^2 C_F 
\nu^{4-D}\int \frac{d^dk}{(2\pi)^D}    \frac{a\pi m_D^2}{k(k^2+m_D^2)^2} \log \frac{k}{T}\,.
\ee
This expression is still UV divergent in $k$. This is due to the kinetic term $-k^2/2m$ in $a$. We expect this divergence to be cancelled by the IR contribution at the scale $k\sim \mu$ of  \Eq{deltam}. In this case we must take the full one-loop longitudinal gluon propagator, but we may treat the selfenergy $\Pi$ as a perturbation. We get,
\be
\delta m\vert_{k\sim \mu}  \simeq  -g^2 C_F 
\nu^{4-D}\int \frac{d^dk}{(2\pi)^d}  \int_0^\infty \frac{dk^0}{2\pi} \Big(\frac{a}{k_0^2} \Big) N_B(k^0)  \Big[ -i\frac{\Pi^R(k_0,k)-\Pi^A(k_0,k)}{k^4} \Big] \,,
\ee
where $a\simeq -k^2/2m$ since for $k\sim \mu$, $\mu^2/m \gg E$. In the region $k_0\sim k\sim \mu$, $k_0 \gg a$ and the expression above follows from expanding the denominators in \Eq{deltam} in $a$. In the region $k_0\sim T \ll m_D\sim \mu^2/m$ the expression above is not correct in general. However, it has the same UV behavior in $k_0$ as \Eq{deltam}, and this is enough to extract the right $\log T$ behavior.
In the region $k_0\sim T$, we may use $k_0\ll k$ to simplify $\Pi$. The region $k_0\sim k$ is independent of $T$ since we can approximate $N(k_0)\sim 1$ and we only need it to cancel the $1/\lambda$ pole from the UV divergence of the $k_0\sim T$ region.
Then we only need the $k_0\to 0$ behavior of the $k_0\sim k$ region and hence we can also use $k_0\ll k$ to simplify $\Pi$. We then have,
\bea
\Pi^R(k_0,k)-\Pi^A(k_0,k) &\simeq &\frac{i\pi m_D^2}{4\mu^2 k}\Big[\left(2k_0\mu^2+k^2\mu-\frac{k^3}{3}-\frac{k^2k_0}{2}\right) \theta(2\mu-k-k_0) \nn\\ 
&& +\left(2k_0\mu^2-k^2\mu+\frac{k^3}{3}-\frac{k^2k_0}{2}\right) \theta(2\mu-k+k_0)+{\cal O}(k_0^2k,k_0^2\mu)\Big] \,.
\eea
In the $k_0\sim T$ contribution we may simply drop $k_0$ from the $\theta$ functions, and recover, 
\be
\Pi^R(k_0,k)-\Pi^A(k_0,k)\simeq \frac{i\pi m_D^2 k_0}{ k}\left( \left(1-\frac{k^2}{4\mu^2}\right) \theta(2\mu-k)+{\cal O}(k_0k,k_0\mu)\right) \,.
\ee
In the $k_0\sim k$ contribution, however, we need to keep $k_0$ in the $\theta$ functions in order to avoid scaleless integrals in $k_0$. Putting together the $k_0\sim T$ and the $k_0\sim k$ regions, we obtain,
\be
\label{muIR}
\delta m\vert_{k\sim \mu}  \simeq  \frac{ g^2 C_F\pi m_D^2}{2m}\frac{\nu^{4-D}\Omega_d}{(2\pi)^D}\int_0^{2\mu}dkk^{d-4} \left(1-\frac{k^2}{4\mu^2}\right)\left[\gamma +\log\frac{2\mu-k}{2\pi T}\right]\,.
\ee
Finally, putting together \Eq{mDUV} and \Eq{muIR} we get for the $T$-dependent log-enhanced contributions to the mass shift,
\be
\delta m=\frac{\als C_F m_D^2}{2\pi m}\log \frac{m_D}{T}\left(\log \frac{\mu}{m_D} + {\cal O}(1)\right)\,.
\ee
The scale $m_D$ in the $\log m_D/T$ above is arbitrary. It can be replaced by any other scale since we have not calculated the $T$-independent pieces at this order because they are subleading.

\end{document}